%% file: paper.tex
\pdfoutput=1

\documentclass[iop, apjl, twocolappendix, numberedappendix]{emulateapj}

\usepackage{xspace}
\usepackage{amsmath}
\usepackage{framed} 
\usepackage{txfonts}
\usepackage{epstopdf}
\usepackage{color}
\usepackage{rotating}
\usepackage{natbib}
\usepackage{ulem}
\usepackage{xspace}
\usepackage[colorlinks=true,urlcolor=blue,linkcolor=blue,citecolor=blue]{hyperref}

\newdimen\hssize
\newdimen\hdsize
\input{commands}

\input{macros}
\input{citation_fix}

\shorttitle{Splashback radius and Halo Assembly bias}
\shortauthors{More, S. et al.}

\slugcomment{To be submitted to the Astrophysical Journal}

\begin{document}


\def\figdir{.}
\def\figext{pdf}


\title{Detection of the Splashback Radius and Halo Assembly bias of Massive Galaxy Clusters}
\author{
Surhud More \altaffilmark{1}, 
Hironao Miyatake \altaffilmark{2,3,1}, 
Masahiro Takada \altaffilmark{1}, 
Benedikt Diemer \altaffilmark{4}, 
Andrey V. Kravtsov\altaffilmark{5,6,7},
Neal K. Dalal \altaffilmark{8, 1},
Anupreeta More \altaffilmark{1},
Ryoma Murata \altaffilmark{1,9},
Rachel Mandelbaum\altaffilmark{10},
Eduardo Rozo\altaffilmark{11},
Eli S. Rykoff\altaffilmark{12},
Masamune Oguri\altaffilmark{9,13,1},
David N. Spergel\altaffilmark{3,1}
}

\affil{
$^1$ Kavli Institute for the Physics and Mathematics of the Universe (WPI),
Tokyo Institutes for Advanced Study, The University of Tokyo,\\ 5-1-5 Kashiwanoha, Kashiwa-shi, Chiba, 277-8583, Japan;{\tt surhud.more@ipmu.jp}\\
$^2$ Jet Propulsion Laboratory, California Institute of Technology, Pasadena, CA 91109, USA\\
$^3$ Department of Astrophysical Sciences, Princeton University,
 Peyton Hall, Princeton NJ 08544 USA\\
$^4$ Harvard-Smithsonian Center for Astrophysics, 60 Garden St., Cambridge, MA 02138 USA \\
$^5$ Department of Astronomy and Astrophysics, The University of Chicago, Chicago, IL 60637 USA \\
$^6$ Kavli Institute for Cosmological Physics, The University of Chicago, Chicago, IL 60637 USA \\
$^7$ Enrico Fermi Institute, The University of Chicago, Chicago, IL 60637 USA \\
$^8$ Department of Physics, University of Illinois Urbana-Champagne,
1110 West Green Street Urbana, IL 61801-3080 USA\\
$^{9}$ Department of Physics, University of Tokyo, 7-3-1 Hongo, Bunkyo-ku, Tokyo 113-0033 Japan\\
$^{10}$ McWilliams Center for Cosmology, Department of Physics, Carnegie
Mellon University,
 Pittsburgh, PA 15213 USA\\
$^{11}$ Department of Physics, University of Arizona, 1118 E 4th St,
Tucson, AZ 85721 USA \\
$^{12}$ SLAC National Accelerator Laboratory, Menlo Park, CA 94025  USA \\
$^{13}$ Research Center for the Early Universe, University of Tokyo, 7-3-1 Hongo,
Bunkyo-ku, Tokyo 113-0033, Japan \\
}


\begin{abstract}
We show that the projected number density profiles of SDSS photometric galaxies
around galaxy clusters displays strong evidence for the splashback radius, a
sharp halo edge corresponding to the location of the first orbital apocenter of
satellite galaxies after their infall. We split the clusters into two subsamples
with different mean projected radial distances of their members, $\ave{\rmem}$,
at fixed richness and redshift, and show that the sample with smaller
$\ave{\rmem}$ has a smaller ratio of the splashback radius to the traditional
halo boundary $\rtom$, than the subsample with larger $\ave{\rmem}$, indicative
of different mass accretion rates for the two subsamples. The same cluster
samples were recently used by Miyatake et al.~to show that their large-scale
clustering differs despite their similar weak lensing masses, demonstrating
strong evidence for halo assembly bias. We expand on this result by presenting a
6.6-$\sigma$ detection of halo assembly bias using the cluster-photometric
galaxy cross-correlations. Our measured splashback radii are smaller, while the
strength of the assembly bias signal is stronger, than expectations from N-body
simulations based on the $\Lambda$-dominated, cold dark matter structure
formation model. Dynamical friction or cluster-finding systematics such as
miscentering or projection effects are not likely to be the sole source of these
discrepancies. 
\end{abstract}



\section{Introduction}
\label{sec:intro}

Dark matter halos with masses larger than $10^{14}\msunh$ collapse out of dense
peaks in the primordial Gaussian density fluctuations that are believed to
originate from quantum fluctuations in cosmic inflation \citep[see
e.g.,][see \citealt{Kravtsov:2012} for a recent review]{Kaiser:1984,bardeen_etal86}. Clusters of galaxies form
within such massive dark matter halos. The large scale
clustering amplitude of the halos hosting galaxy clusters is thus heavily biased
compared to the underlying matter distribution
\citep{Kaiser:1984,Mo:1996,Sheth:2001,Tinker:2010}. 

Although the large scale clustering amplitude of dark matter halos is primarily
expected to depend upon the height of the initial density peak out of which a
halo collapses (therefore its halo mass),
it can have secondary dependencies on other parameters related
to the assembly history of the halo, such as the radial profile of the initial
peak, especially on cluster scales \citep{Dalal:2008}. The dependence of the large scale clustering amplitude on parameters
other than the halo mass has been broadly referred to as halo assembly bias, and
has been studied in great detail using cosmological simulations
\citep{Sheth:2004, Gao:2005,*Gao:2007, Wechsler:2006, Li:2008}.

Halo assembly bias has however been difficult to establish in astrophysical
observations. A clean detection of halo assembly bias requires identifying
samples of isolated halos which are matched in their halo masses but differ
in their assembly histories. There have been several claims of detection of halo
assembly bias on galaxy scales in the literature \citep[e.g.,][]{Yang:2006,
Tinker:2012, Hearin:2014}. However, \citet{Linetal:2015} investigated the first of
these claims and found no strong evidence for halo assembly bias on galaxy scales. The
difference in the conclusions was a result of contamination of the halo
samples by satellite galaxies, or the differences in halo masses of the samples
used to look for halo assembly bias \citep{Linetal:2015}. 

Recently, \citet{Miyatake:2016}
presented the first evidence of halo assembly bias on cluster scales. Galaxy
clusters offer two advantages: first, the probability of a cluster-sized halo
being a satellite of an even bigger halo is much smaller than in the case of
galaxies, and secondly, the weak gravitational lensing signal can be used to
match galaxy cluster subsamples for their halo masses with a greater accuracy. 

The galaxy cluster subsamples used by \citet{Miyatake:2016}
were drawn from the SDSS \redms galaxy cluster catalog, and were matched in redshift
and richness distribution, but differed in the compactness of the member galaxy
distribution. These samples were shown to have very similar masses based on weak
lensing, but had different large scale biases. The main goal of this paper is
to observationally establish the connection between the member galaxy
distribution and the mass assembly of these cluster subsamples without relying
on proxies related to complicated baryonic physics, such as the star formation
rates.

For this purpose, we use a unique probe of the mass assembly of galaxy
clusters, which relies on the observational detection of the edges of galaxy
clusters. Models of self-similar secondary infall of matter onto a spherical
overdensity predict the presence of a density jump at the location where
recently accreted material is reaching its first apocenter, associated with the
last density caustic \citep{fillmore_goldreich84, Bertschinger:1985}.  Although
the collapse of matter onto realistic density peaks in cold dark matter models
is considerably more complex than that envisioned in these models, the last
density caustic manifests itself as a sharp steepening of the density profile in
dark matter halos \citep{Diemer:2014}. 

The location of this density caustic, also called the splashback radius or the
turnaround radius, can be used to define a physical boundary for dark matter
halos \citep{More:2015}.  The splashback radius crucially depends upon the mass
accretion rate of the collapsing halo \citep{vogelsberger_etal11, Diemer:2014,
Adhikari:2014}.  For halos of the same mass, large accretion 
rate results in  a smaller splashback radius. The physical reason is simple: the deeper
the halo potential well gets during the orbit of a dark matter particle, the
smaller is the value of its apocenter.

As discussed in \citet{More:2015}, hints for
the splashback radius may have been seen before for individual clusters
\citep{Rines:2013, Tully:2015, Patej:2015}. In this paper, we will harness the
power of statistics to present the first high signal-to-noise detection of the
splashback radius for our galaxy cluster subsamples. We will use the splashback
radius to establish that these galaxy cluster subsamples have different mass
accretion rates, and have different large scale clustering amplitude, a signature of halo
assembly bias.

The paper is organized as follows: 
\begin{itemize}
\item
        Section~\ref{sec:data} describes the cluster subsamples and the Sloan
        Digital Sky Survey (SDSS) photometric galaxy data which form the basis
        of our study, and the methods we adopt in order to obtain
        the measurements of the galaxy number densities around our cluster
        subsamples. 
\item
       Section~\ref{sec:results} presents our measurements of the galaxy number
       densities around our cluster subsamples, our inferences for the location
       of the splashback radius from these measurements, and our detection of
       halo assembly bias.  
\item
       Section~\ref{sec:sim_compare} presents the predictions for the location
       of the  splashback radius and the amount of halo assembly bias from
       numerical simulations in the context of the standard cosmological model.
       In particular, we discuss a number of systematic effects, which could
       affect our interpretation.
\item
       The broad implications of our results are discussed in
       Section~\ref{sec:discuss}, and conclusions and a summary is presented in
       Section~\ref{sec:summary}.
\end{itemize}

Throughout this paper, we adopt a flat $\Lambda$CDM cosmological model
with matter density parameter $\omm=0.27$ and the Hubble parameter
$h=0.7$, unless otherwise stated. We use $\log$ to denote logarithms with
respect to base 10. We will use $r$ to denote three dimensional distances, and
$R$ for projected distances between galaxies or between galaxies and cluster
centers. For cases where we want to preserve notations from previous papers,
such as using $\rsp$ for the splashback radius, we will specifically mention
$2$d or $3$d to avoid confusion. The subscript ${\rm 200m}$ on halo mass $M$ or
radius $R$ will refer to the mass or radius corresponding to spherical
overdensity halos such that their boundaries enclose 200 times the mean matter
density of the Universe.

\section{Data and Methods}
\label{sec:data}

We start from the publicly available catalog of galaxy clusters
identified from the SDSS DR8 photometric galaxy catalog by the {\it
red}-sequence {\it Ma}tched-filter {\it P}robabilistic {\it Per}colation
(\redm) cluster finding algorithm \citep[v5.10, see the
website\footnote{\url{http://risa.stanford.edu/redmapper/}} for details
and ][]{Rykoff:2014,Rozoetal:2014}. The cluster finder uses the
$ugriz$ magnitudes and their errors, to identify overdensities of
red-sequence galaxies with similar colors as galaxy clusters. For each
cluster, the catalog contains an optical richness estimate $\lambda$, a
photometric redshift estimate $z_\lambda$, as well as the position and
centering probabilities of 5 candidate central galaxies ${\pcen}$. A separate
member galaxy catalog provides a list of members for each cluster, each
of which is assigned a membership probability, ${\pmem}$.

The parent cluster catalog used in \citet{Miyatake:2016} consists of an
approximately volume limited sample of $8,648$ \redms clusters with
$20<\lambda<100$ and $0.1\le z_{\lambda}\le 0.33$. The average and the median redshift of our
subsamples are $0.24$ and $0.25$, respectively. Throughout this paper we use
the position of the most probable central galaxy in each cluster region as a
proxy of the cluster center. However, we will discuss the effect of miscentering
on our conclusions in Section~\ref{sec:background}.

In this paper, we subdivide this galaxy cluster sample into two subsamples
following the same procedure as in \citet{Miyatake:2016}.  Briefly, we obtain
the average projected cluster-centric separation of member galaxies,
$\ave{\rmem}$, for each cluster, and compute the median $\ave{\rmem}$ as a
function of richness and redshift \footnote{While computing the average, we weight each
galaxy's cluster centric distance with its membership probability ($\pmem$).}. We
use this median to divide the parent sample into two subsamples. The large- and
small-$\ave{\rmem}$ subsamples, labelled as low- and high-${\cgal}$,
respectively, in this paper, consist of 4,235 and 4,413 clusters, respectively.  

In order to compute galaxy surface number density around these cluster
subsamples, we make use of the photometric galaxy catalog from SDSS DR8
\citep{Aihara:2011}. We exclude galaxies with any of the following flags: {\sc
saturated, satur$\_$center, bright, deblended$\_$as$\_$moving}. We correct the
magnitudes for galactic dust extinction using the maps of \citet{Schlegel:1998},
and use all photometric galaxies with extinction corrected $i$-band model
magnitudes brighter than $21.0$ and with magnitude errors less than $0.1$. 

We compute the stacked surface number density of the SDSS photometric galaxy
samples around each of our cluster subsamples as a function of comoving
projected separation, $R$, from the galaxy cluster center. Since our cluster
subsamples span a wide range in redshift ($0.1\le z\le 0.33$), the surface
density profiles around lower redshift clusters will systematically contribute
galaxies from a fainter photometric galaxy population. To avoid such biases, for
our fiducial analysis, we only count cluster-galaxy pairs if the photometric
galaxy has an absolute magnitude of $M_i-5\log h<-19.43$\footnote{Note that we
do not use any k-corrections or corrections for luminosity evolution here,
since the redshifts of the photometric galaxies are quite uncertain.}, assuming
that it is located at the redshift of the cluster (this limit corresponds to an
apparent magnitude of $m_i=21$ at $z=0.33$ for our assumed cosmological model).
Additionally, we will also present results for photometric galaxies that are one
and two magnitudes brighter than our fiducial measurement, to explore the
dependence of the splashback radius on the magnitude of photometric galaxies
used.

We expect that the surface density measurement will consist of galaxies
correlated with the galaxy clusters under consideration as well as uncorrelated
galaxies in the foreground and the background. To determine this uncorrelated
component, we compute the galaxy surface density around a sample of random points. We use 
100 times larger number of random points than the number of clusters in our subsamples \footnote{We have tested that the
use of the improved random catalog from \citet{Rykoff:2016} does not change any
conclusions in this paper.}. These random points incorporate
the survey geometry, depth variations, and distributions of clusters in redshift
and richness. We subtract the background around random points from the total to
obtain the surface density of correlated galaxies, $\sigmag(R)$. We use 102
jackknife regions in order to compute the covariance in the measurements of
$\sigmag(R)$ with typical size of $10\times10$ sq.~deg. which corresponds to
about $100\times100 ~(h^{-1}{\rm Mpc})^2$ at the median redshift of our cluster
subsamples. The jackknife regions are thus significantly larger compared to the
scales of interest in this paper, justifying the assumptions behind the
jackknife errors.

Using the measurement of the galaxy surface densities, we would like to infer
the location of the splashback radius of our cluster subsamples, i.e., the
steepest logarithmic slope of the galaxy number density distributions in three
dimensions. Given that the splashback radius is expected to be of the order of $\rtom$ of
our halos, we fit the surface densities in the range $[0.1,~5.0]\mpch$.  The
location of the steepening in three dimensions is expected to be different from
that in projection \citep{Diemer:2014}.  Therefore, we will use a 3-dimensional
parameterization first proposed by \citet{Diemer:2014} to forward model the
projected measurements \citep[see also][]{More:2015}.  The model consists of
inner and outer surface density profiles with a smooth transition between the
two,
\begin{eqnarray}
        \rhog(r) &=& \rhoginner {\ftrans} +
        \rhogouter\,, \nonumber\\ 
        \rhoginner &=& \rho_\rms \exp\left( -\frac{2}{\alpha}\left[
        \left(\frac{r}{\rs}\right)^\alpha - 1\right]\right)\,, \nonumber\\
        \rhogouter &=& \rho_\rmo \left( \frac{r}{\rout}
\right)^{-\se}\,, \nonumber \\
        {\ftrans} &=& \left[1 +
		       \left(r/\rt\right)^{\beta}\right]^{-\gamma/\beta},
        \nonumber \\
        {\sigmag}(R) &=& 2 \int_{0}^{z_{\rm max}}
	 \rhog\!\left(\sqrt{R^2+z^2}\right) dz~.
	 \label{eq:model_profile}
\end{eqnarray}
Note that the above fitting formula, which is an Einasto profile in the inner
regions which transitions to a power law in the outer regions, is able to reproduce the dark matter
profile around halos, and is flexible enough to reproduce the simulation results
compared to other fitting formulae such as the Navarro-Frenk-White model and the
halo model \citep{Navarro:1996ApJ,OguriHamana:2011,Hikage:2013}. Here we simply
assume that the same fitting formula is also flexible enough to reproduce the
galaxy surface density measured from the SDSS data.\footnote{Tests using subhalo
density profiles around cluster scale halos from simulations presented in
Appendix~\ref{app:fittest}, as well as the reasonable values of $\chi^2$ values
we obtain for describing the observed measurements justify this choice.} We
chose the maximum projection length $z_{\rm max}=40\mpch$ as our default value,
and we have checked that the location of the splashback radius is insensitive to
this choice, in particular reducing $z_{\rm max}$ to be even as small as
$10\mpch$.

\begin{figure*} 
\centering{
\includegraphics[scale=1.0]{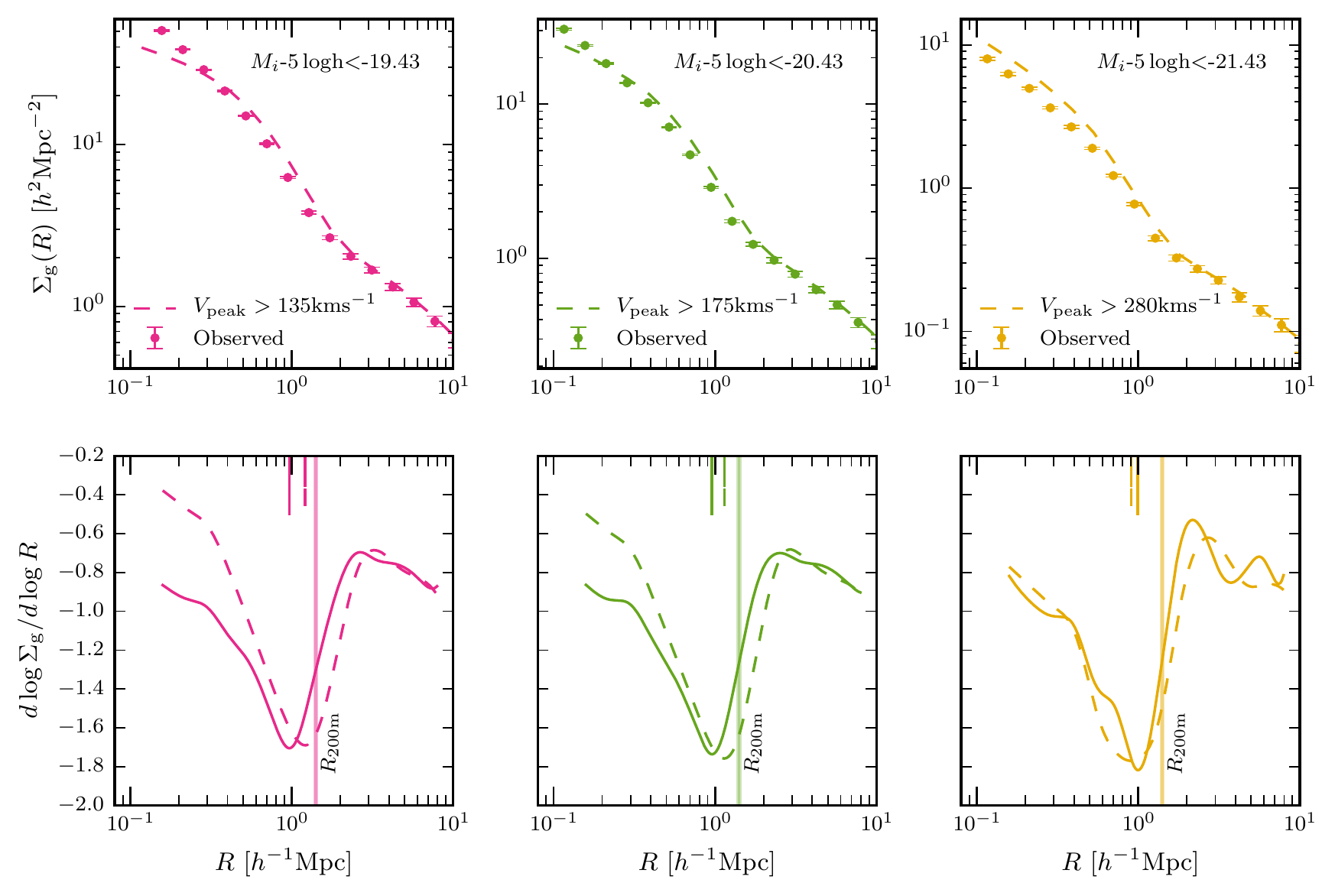}}
\caption{
        {\it Top panels}: The surface number density profiles, $\sigmag(R)$, of
        SDSS photometric galaxies with different magnitude thresholds around the
        entire \redms cluster sample with $z\in[0.1, 0.33]$ and richness
        $\lambda\in[20, 100]$, are shown using symbols with errorbars. The
        dashed lines correspond to (sub)-halo surface density profiles in the
        Multidark Planck II simulation around clusters with the mass threshold
        similar to our sample at $z=0.248$. The threshold on subhalo $V_{\rm
        peak}$ values roughly correspond to the magnitude thresholds in each
        panel and were obtained by subhalo abundance matching (see Appendix
        ~\ref{app:amatch}). {\it Bottom panels}: The logarithmic slope of the
        surface density profiles are shown using solid and dashed lines for the
        observed galaxy and the subhalo surface density profiles, respectively.
        The observed slope of the surface density profile has a shape which is
        similar to that expected from simulations. Note that although the
        surface density profiles both in observations and simulations exhibit
        similar steepening, the corresponding radii of the steepest slope are at
        slightly different locations.
}
\label{fig:obs_all}
\end{figure*}

\begin{figure*} 
\centering{
\includegraphics[scale=1.0]{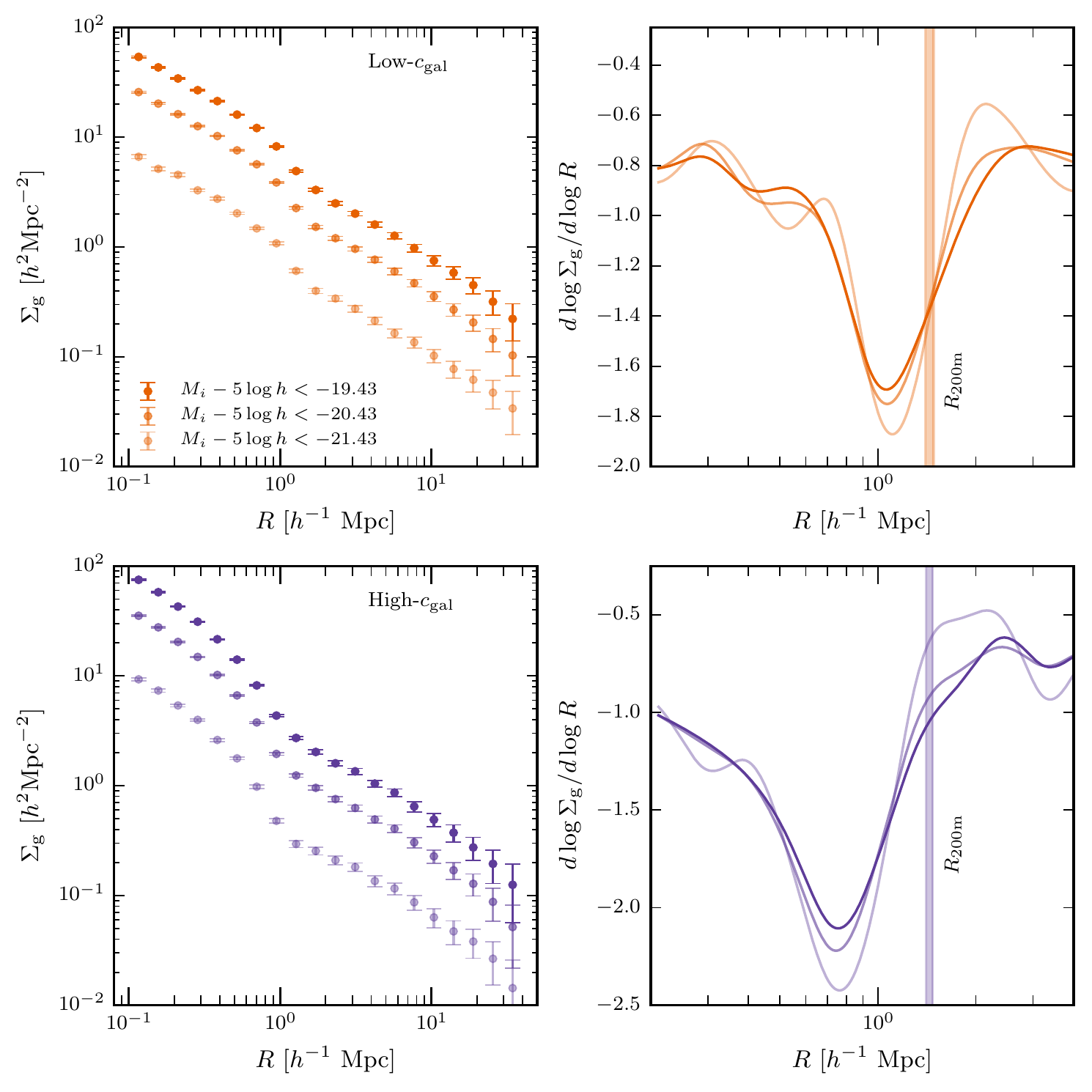}}
\caption{
        {\it Left panels}: The surface number density profiles, $\sigmag(R)$, of
        SDSS photometric galaxies around the low-$c_{\rm gal}$ (top) and high-$c_{\rm gal}$ (bottom) cluster subsamples. The three different point types
        with errorbars in each panel correspond to the three different magnitude
        limited samples of photometric galaxies we use. {\it Right panels:} The
        logarithmic density slope of the surface number density profiles
        obtained after smoothing the data points with an improved Savitzky-Golay filter
        (order 3, window size 5). The two-dimensional splashback radius
        corresponds to the location of the steepest slope or the minimum of
        $d\log\sigmag/d\log R$. For comparison, the shaded regions correspond to
        the traditional halo boundary, $R_{\rm 200m}$, estimated using the
        posterior distribution of halo masses from the weak lensing profile for
        each cluster subsample from \citet{Miyatake:2016} (both are consistent
        with each other within the errorbars).  Note that the steepest slope (i.e., the minimum in
        $d\log\sigmag/d\log R$) occurs at different locations for the two cluster
        subsamples.
}
\label{fig:obs}
\end{figure*}

Given that the parameter ${\rout}$ and $\rho_\rmo$ are entirely degenerate with
each other, we fix ${\rout}=1.5 \mpch$. We find that allowing $\alpha$ to vary
freely results in an almost perfect degeneracy between $\rho_\rms$ and $r_\rms$,
with very little impact on the location of the steepening of the galaxy density
profiles. Therefore we use a prior on $\log \alpha = \log 0.2 \pm 0.6$,
centered at the value expected for the dark matter halos corresponding to our
mass estimates from weak lensing \citep{Gao:2008}. For our fiducial modeling
scheme, we also use priors on $\log \beta=\log 4.0\pm0.2$ and $\log
\gamma=\log 6.0\pm0.2$ centered around the values recommended by
\citet{Diemer:2014} and constrain the parameters $\rs$ and $\rt$ to lie within
$[0.1, 5.0]\mpch$. 

For our default modeling scheme, we assume that the most probable central
galaxy for every cluster (one with the highest $\pcen$), assigned
in the \redms catalog, resides at the true center of gravitational potential in
each cluster region. However, as studied in \citet{Miyatake:2016} \citep[see
also][]{Hikage:2013}, some fraction of the central galaxies in our cluster
subsamples may be mis-centered, characterized by offset radii ranging from
$400 \kpch$ possibly up to $800 \kpch$. If such mis-centered clusters are indeed present
in large numbers, our measurements of the splashback radius would be biased
high. We will present tests for the effects of mis-centering  in Section~\ref{sec:background} below.

\renewcommand{\tabcolsep}{0.1cm}
\begin{deluxetable*}{ccccccccccccc}
\tablecaption{Posterior distribution of parameters from 
the MCMC analysis
 \label{tab:post}
}
\startdata
 \hline\hline
\input{Latex_table.tex}
 \enddata
 \tablecomments{
The different rows list the $68\%$ confidence intervals on the model parameters
(see Eq.~\ref{eq:model_profile}) given the surface number density data shown in
Figure~\ref{fig:obs}.  The $\chi^2$ per degree of freedom as well as the
inferred $2$-d and $3$-d splashback radius are also shown in the last three
columns. 
}
\end{deluxetable*}

We will use the affine invariant Markov Chain Monte Carlo sampler of
\citet{Goodman:2010}, as implemented in the software package {\it emcee}
\citep{Foreman-Mackey:2013}, in order to sample from the posterior distribution
of the parameters, $\log\rhos$, $\log\rs$, $\log \alpha$, $\log\rt$,
$\log\gamma$, $\log\beta$, $\log\rho_\rmo$ and $\se$, given the galaxy surface
density measurements and the stated priors.

As a test of our fitting method, in Appendix~\ref{app:fittest} we apply it to
projected number density distributions of (sub)-halos around galaxy clusters in
numerical simulations (see below for the details of the simulations)
and show that we are able to recover the location of the
steepening of the three dimensional density distribution of subhalos quite
accurately with our modelling scheme.
\begin{figure*} 
\centering{
\includegraphics[scale=1.0]{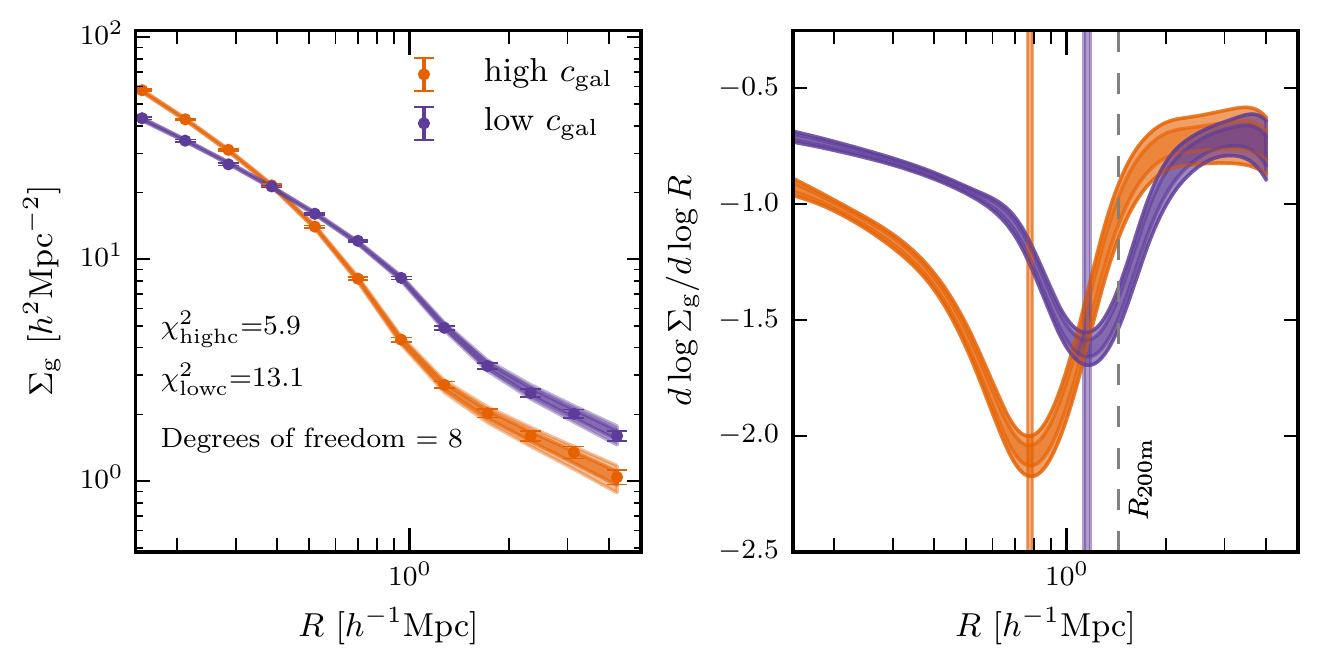}}
\caption{
        The surface number density profiles, $\sigmag(R)$, of our fiducial
        sample of SDSS photometric galaxies around the two cluster subsamples
        are shown in the left hand panel. The shaded regions show the 68 and 95
        percent confidence regions of our model fit to the data. The right
        hand panel shows the inferred constraints on the logarithmic slope
        of $\sigmag(R)$ for the two subsamples. The splashback radius in $2$d,
        $\rsptod$, corresponds to the location of the steepest slope or the minimum of
        $d\log\sigmag/d\log R$. The 68 percent constraints on $\rsptod$ are marked
        with vertical shaded regions. These minima occur at significantly
        different locations for the two cluster subsamples. The traditional halo
        boundary, $\rtom$, is marked by the grey dotted vertical line.
}
\label{fig:fit}
\end{figure*}
\begin{figure} 
\centering{
\includegraphics[scale=1.0]{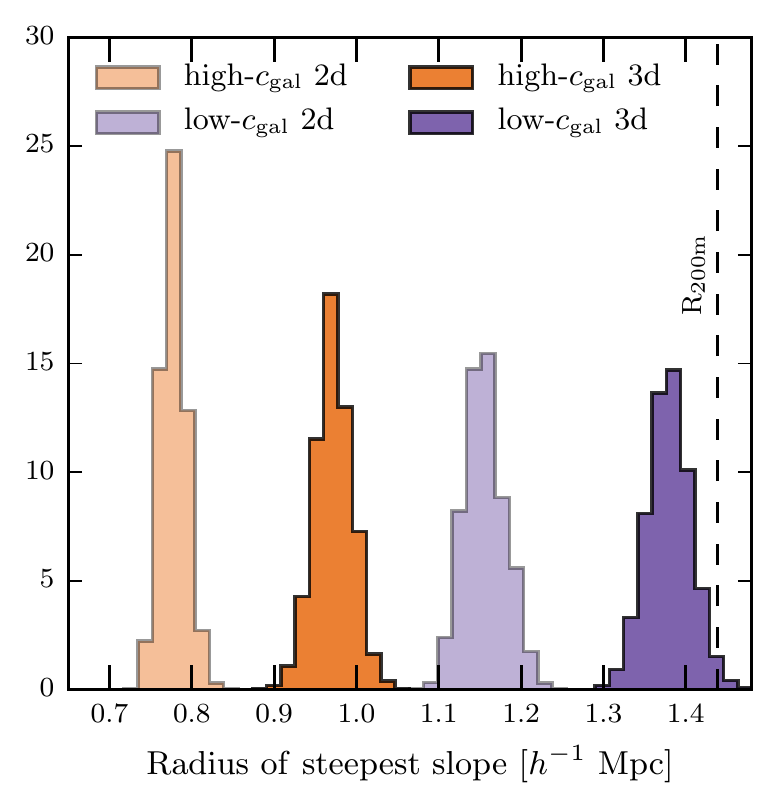}}
\caption{
        The posterior distributions for the location of the steepest slope of
        the galaxy density profiles around the high- and low-$\cgal$ cluster
        subsamples are shown in orange and purple colored histograms,
        respectively.  The light shaded histograms correspond to the location of
        the steepest slopes of the surface density profiles ($2$-d), while the
        dark shaded histograms correspond to the location of steepest
        slope of the $3$-d number density profiles inferred by our fits. The locations of the
        steepest slopes for the two cluster subsamples are significantly
        different, implying a different mass accretion rate onto these cluster
        subsamples.
}
\label{fig:deriv}
\end{figure}
\begin{figure*} 
\centering{
\includegraphics[scale=1.0]{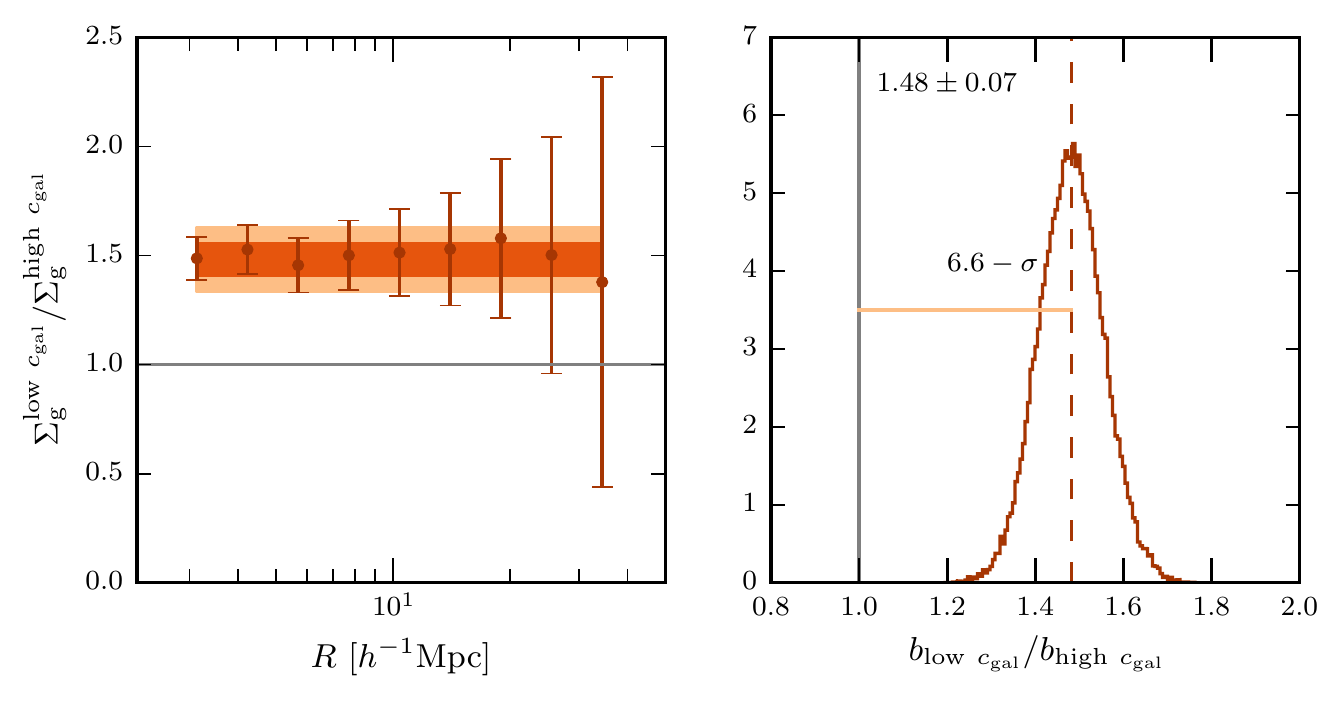}}
\caption{
        Detection of the halo assembly bias. {\it Left panel:}\/ The ratio of the surface
        number density profiles of our fiducial samples of photometric
        galaxies around the two galaxy cluster subsamples. The shaded regions
        correspond to the $1$- and $2$-sigma confidence regions for a single
        constant parameter fit to these data.  {\it Right panel:\/} The posterior
        distribution of the ratio given the measurements shown in the left
        panel. We detect the assembly bias -- difference in the halo biases of the two samples -- at
        $6.6\sigma$. There is a significant covariance in the errors, hence the
        small point-to-point variation given the errors. The quoted significance
        accounts for the covariance.
}
\label{fig:assembly}
\end{figure*}
\begin{figure} 
\centering{
\includegraphics[scale=1.0]{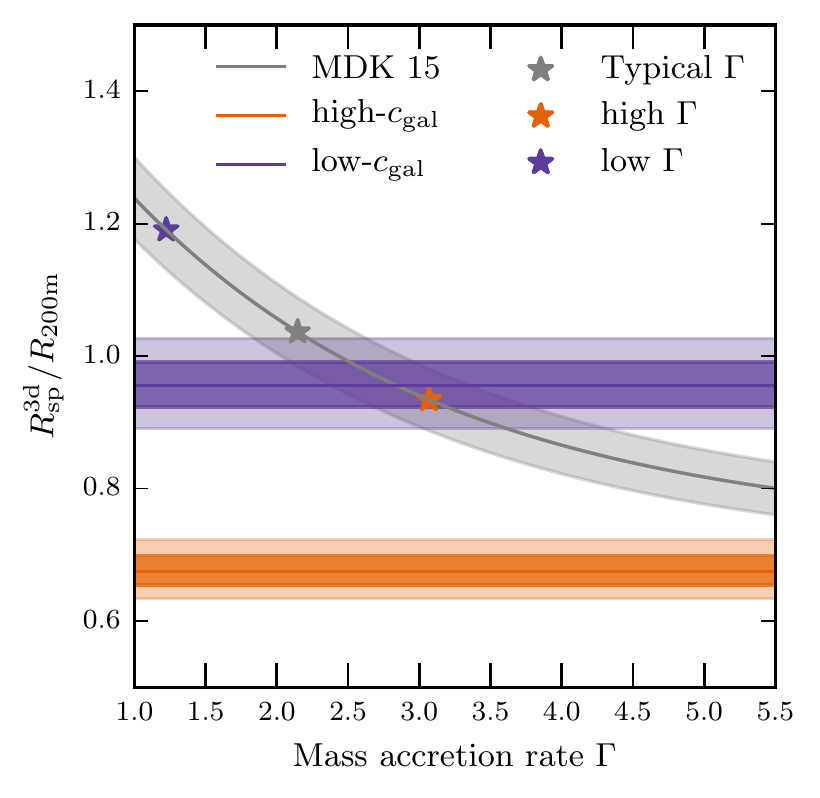}}
\caption{
The dependence of $\rsp/\rtom$ on the accretion rate $\Gamma$ at $z=0.24$ predicted by $\Lambda$CDM, shown by a gray line with $5\%$ uncertainty \citep{More:2015} . The dark and
faint shaded purple (orange) regions display the 68 and 95 percent confidence
limits on the splashback radius in 3d for our low- (high-) ${\cgal}$ sample. The gray, orange and purple stars 
correspond to the typical splashback radii for a typical accretion rate, as well as slow and fast accreting clusters from numerical simulations (see
 text for details). The observed
values of the splashback radii are significantly smaller from the predicted values their halo masses, even if we consider halos with typical
accretion rates.
} 
\label{fig:derivmdk}
\end{figure}
\begin{figure*} 
\centering{
\includegraphics[scale=1.0]{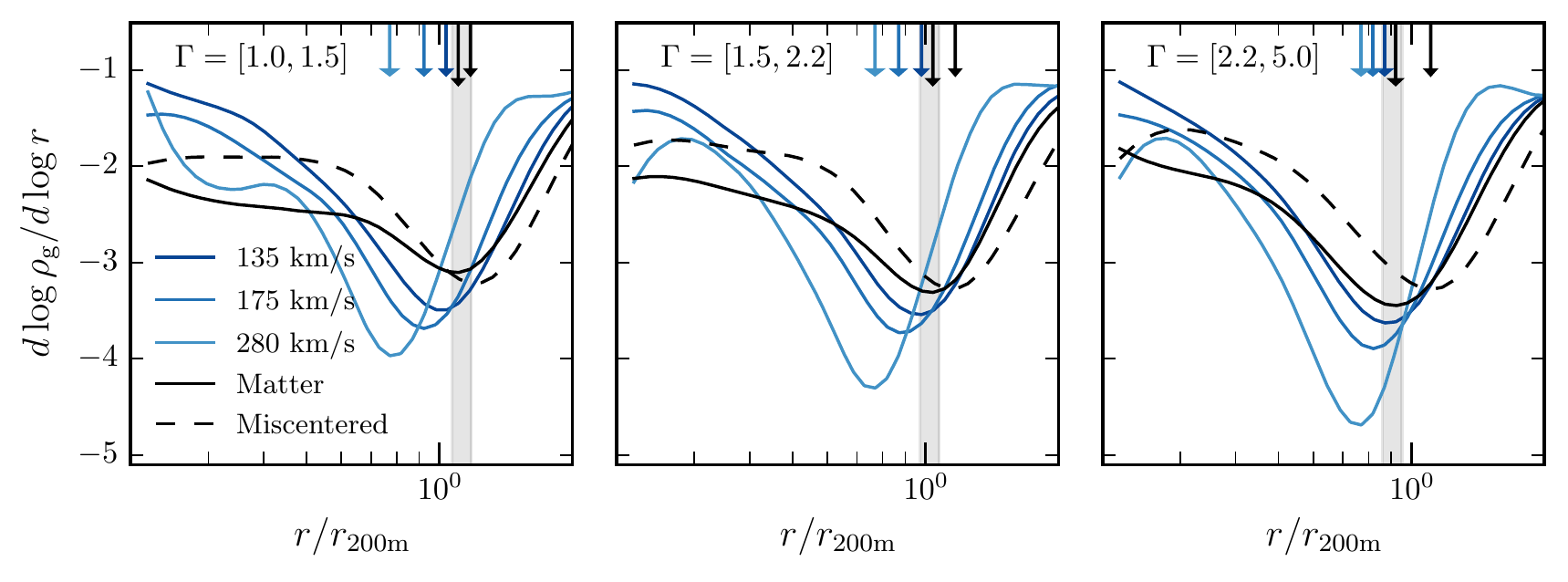}}
\caption{
Comparison between the logarithmic slope of the density profile for matter and
that of subhalos selected using different ${\vpeak}$ thresholds as indicated in
the legend. Different panels correspond to halos with different mass accretion
rates, $\Gamma$. The arrows indicate the location of the steepest slope, or the
splashback radii for the corresponding population, with the longest arrow used
to represent dark matter. The grey vertical bands corresponds to the fitting
function for $\rsp/\rtom$ similar to the one by \citet{More:2015}, but for the
mean profiles, and includes an uncertainty of $\pm5\%$.
}
\label{fig:sh_matter_rsp_3d}
\end{figure*}

\section{Results}
\label{sec:results}
\subsection{Splashback in galaxy number density profiles}
\label{sec:splashback}

We begin by presenting how the stacked surface density profile of galaxies,
$\sigmag(R)$, around the entire parent sample of \redms clusters, described in
the previous section, varies with galaxy samples of different absolute magnitude
thresholds. These measurements are shown in the
top panels of Figure~\ref{fig:obs_all} using points with errorbars. The
brightness of the photometric galaxy sample increases from left to right. The
solid lines in the bottom panels correspond to the profiles of the logarithmic
slope of the galaxy surface densities. These slope profiles were obtained by
using the Savitzky-Golay algorithm to smooth the observed measurements, by
fitting a third-order polynomial over a window of five neighbouring points, and
then using a cubic spline to interpolate between these smoothed measurements. In
contrast with the traditional Savitzky-Golay algorithm, we explicitly account
for the covariant errors on these data points, as proposed by
\citet{More:2016a}\footnote{\url{https://github.com/surhudm/savitzky\_golay\_with\_errors}}.

In Appendix~\ref{app:amatch}, we have used subhalo abundance matching to obtain
an estimate of the approximate ${\vpeak}$ (the maximum circular velocity of a
halo throughout its entire history) value of dark matter subhalos\footnote{Our
use of the term subhalos henceforth includes isolated host halos as well, not
just satellite halos. We will use the term halos explicitly when referring to
only isolated halos.} hosting our photometric galaxies as a function of their
magnitude. To compare the observed surface density profiles with those expected
from the standard structure formation model, we utilize Multidark Planck II
(MDPL2), a $3840^3$ particle cosmological N-body simulation with box size of
$1h^{-1} {\rm Gpc}$ and mass resolution of $1.51\times 10^9h^{-1} M_\odot$
\citep{Klypin:2014}. We also use the associated halo catalogs found using
Rockstar, a halo finder which groups particles into halos using their phase
space information \citep{Behroozi:2013}\footnote{These catalogs are publicly
available at the website \url{http://www.cosmosim.org}}.

The dashed lines in the top and bottom panels of Figure~\ref{fig:obs_all}
correspond to the expected subhalo surface density profiles around clusters in
the cosmological simulation Multidark Planck II  at $z=0.248$. We have selected
cluster sized halos above a mass threshold of $10^{14}\msunh$, which results in the same
average halo mass as that of our sample. We have normalized the surface density
profiles in the top panels to match the observations at $\sim11\mpch$. There are
marked similarities between the density profiles of subhalos in simulations and
the galaxies in observations. The surface number densities strongly deviate from
a simple power law and show a clear break on scales of $\sim1\mpch$ in both
observations and simulations. This is most clearly seen in the bottom panels,
where we see that the profiles reach their steepest slope on scales of
$\sim1\mpch$. This steepening, associated with the splashback radius, is also
seen in observations, as expected from subhalo surface density profiles in
simulations. However, it is also clear that the locations where the surface
density profiles reach their steepest slopes are different between observations
and simulations, especially for the left and the middle panels. This discrepancy
between the observed and expected splashback radii is also seen for the cluster
subsamples, which we investigate at length next. We will extensively
quantify, comment and explore this discrepancy in the location of splashback
radius around these cluster subsamples.

The surface density of photometric galaxies around the low- and high-$\cgal$
cluster subsamples are shown in the upper and lower panels of
Figure~\ref{fig:obs} using orange and purple symbols with errorbars,
respectively. The lighter shades correspond to photometric galaxies with
brighter magnitude limits. The solid lines in the upper and lower right hand
panels of Figure~\ref{fig:obs} show the logarithmic slope of the surface density
profiles around the two subsamples. The slopes for both cluster subsamples
reach values steeper than $\sim-1.6$ on either side of $\sim 1 \mpch$.  The
surface density of galaxies around the high-${\cgal}$ cluster subsample reaches
its steepest slope at a smaller radius compared to the low-${\cgal}$ subsample.
The value of the steepest slope is considerably larger for the high-${\cgal}$
cluster subsample than the low-${\cgal}$ subsample. A comparison between the
profiles of galaxies as a function of different magnitude thresholds around any
given cluster subsamples shows very little difference in the location of the
steepest slope in projection.

We fit the galaxy surface density profiles with the model described in the
previous section. The median and the 68 percent confidence intervals of the
posteriors of each of these parameters, as well as the best fit $\chi^2$ values
are listed in Table~\ref{tab:post}. The number of degrees of freedom for our
model is $8$.

We show the 68 and 95 percent confidence regions from the fits to the surface
density of the fiducial sample of photometric galaxies around both our cluster
subsamples in the left hand panel of Figure~\ref{fig:fit}. The corresponding
confidence regions for the logarithmic slope, including marginalization over
other model parameters, are shown in the right hand panel. We use the samples
from the posterior of the model parameter space to infer the location of the
steepest slope of the projected galaxy density profile, $\rsptod$, and its
uncertainty. These numbers are reported for all of our subsamples and for the
different models in Table~\ref{tab:post} as well. 

The location of the splashback
radius can be compared with the traditional halo boundary definition, $\rtom$
for each subsample. This is shown by the vertical shaded bands in the right
panels of Figure~\ref{fig:obs}, as estimated from the posterior distribution of
the halo masses for our two subsamples inferred from the weak lensing
measurement in \citet{Miyatake:2016}. 

We now use the samples from the posterior distribution of model parameters to
infer the constraints on the location of the minimum of the logarithmic
derivative of the three dimensional galaxy density profile, $d\log\rho_\rmg/d\log r$.
The resultant constraints on $\rspthd$ are reported in the penultimate column
of Table~\ref{tab:post}.  The inferred value of $\rspthd$ is always larger than
the corresponding $\rsptod$ for all photometric galaxy samples around both cluster
subsamples, as shown explicitly in Figure~\ref{fig:deriv}. The vertical dashed
line corresponds to the traditional halo boundary definition, $\rtom$, for the
two subsamples.

Note that, for our model, a transition function $\ftrans=1$, would correspond to
a simple density profile: a sum of Einasto profile which describes well the 
inner regions and a power law  profile for the outer
regions. However, the data strongly disfavor such a model, with $\chi^2$ values
ranging from $60$ to $140$ for $9$ degrees of freedom depending upon the cluster
subsample and the photometric galaxies under consideration \footnote{There is
only 1 additional degree of freedom for these models, as we lose only only one
free parameter $r_{\rm t}$, the other parameters $\gamma$ and $\beta$ have
priors in the fiducial modeling scheme and thus do not change the degrees of
freedom.}. Therefore, {\it our measurements imply a steepening of the number
density profile of galaxies around both of our cluster subsamples beyond that
predicted by the Einasto profile}.

\subsection{Detection of Halo Assembly bias}
\label{sec:assembly_bias}

The mean number density profile of galaxies correlated with clusters at large separations
is proportional to the product of the biases of clusters and galaxies in the
photometric sample. We have shown above that these profiles are different for
the low- and high-$\cgal$ cluster subsamples. Given that our cluster samples
have the same redshift distribution, the bias of photometric galaxies should
cancel out in the ratio of the number density profiles, and we can use the ratio
to test whether the two cluster subsamples have different intrinsic clustering
biases, as recently reported by \citet{Miyatake:2016}.

In the left hand panel of Figure~\ref{fig:assembly}, we show the ratio of the
number density profiles for our fiducial sample of photometric galaxies around the
two subsamples of galaxy clusters on scales of $3-40 \mpch$. We fit a constant
ratio to these measurements accounting for the covariance determined from the
jackknife technique. This assumes that the two samples have similar scale
dependence for their bias, and the data support this assumption.
The posterior distribution of the constant ratio obtained
using this procedure is shown in the right hand panel of
Figure~\ref{fig:assembly}. We find a $6.6\sigma$ deviation of the ratio of the
two surface density profiles from unity: $1.48\pm 0.07$. We have thus detected halo assembly
bias -- the two cluster subsamples have the same halo mass based on weak
lensing, but a different large scale halo bias. For comparison, the difference
in the bias ratio that was obtained in \citet{Miyatake:2016} using the weak
lensing signal was $1.64^{+0.31}_{-0.26}$, and that from the auto-correlation
function of galaxy clusters was $1.40\pm0.09$.  The three different measurements
give results which are statistically consistent with each other.

\section{Comparisons with expectations from $\Lambda$CDM model}
\label{sec:sim_compare}

Using the projected galaxy number density profiles around two cluster subsamples
from the \redms catalog, we have shown that these two subsamples have different
profiles, splashback radii, as well as a different clustering bias. We now
compare these measurements to the predictions of the concordance cosmological
$\Lambda$CDM model.

\subsection{Is the splashback radius for the two subsamples at the expected location?}

Using the weak lensing inferred masses for our cluster subsamples, we can
compute the baseline expectation for the location of the splashback radius in
the standard cosmological model. 
The ratio $\rspthd/\rtom$ is expected to depend upon
the accretion rate of the halos as well as redshift \citep{Diemer:2014,
Adhikari:2014, More:2015}. To compute the mean accretion rate on to halos, we
use halos from MDPL2 at $z=0.248$, closest to the median redshift of our \redms subsamples
and select all halos above a certain halo mass threshold\footnote{Using a
cosmological simulation with $\omm=0.27$, we get similar numbers for the
expected mass accretion rates.}. We choose the halo mass threshold such that the
average halo mass of the sample is consistent with the ${\mtom}$ of the \redms
subsamples obtained by \citet{Miyatake:2016}. As a best case expectation, we
divide the halo sample into two based on the dependence of the halo mass
accretion rate on halo mass, $\Gamma({\mtom})=\Delta\log \mvir/\Delta\log a$.
The derivative for $\Gamma$ is computed using a finite difference scheme using
the virial masses at redshift $0.248$ and $0.748$ \citep{Diemer:2014,
More:2015}\footnote{There is very little difference in the average value of
$\Gamma$ if we use a halo mass sample with a threshold on $\nsat$}.

In Figure~\ref{fig:derivmdk}, we compare the location of the splashback radius
with respect to $\rtom$ observed for our cluster subsamples, to the best case
expectations implied by these accretion rates. The grey band corresponds to the
fitting function
\begin{equation}
        \frac{\rspthd}{\rtom} = 0.58
        \left[1+0.63~\omm(z)\right]\left(1+1.08\exp\left[-\frac{\Gamma}{2.26}\right]\right)
        \label{eq:meanMDK15}
\end{equation}
with a 5 percent uncertainty. This fitting function is a good fit to the
splashback radii of dark matter halos in simulations used in
\citet{More:2015}\footnote{The fitting function was calibrated in the redshift
range [0, 4].}, but corresponds to the mean profiles instead of the median. The grey star
corresponds to the typical expected value of $\Gamma$ for halos in the sample,
estimated from the simulations, while the orange and purple stars similarly
correspond to the average $\Gamma$ for the best-case simulation subsamples with
the fastest and slowest accretion rates (see above). The data seem to prefer a
much smaller splashback radius for each of our cluster subsamples
($\rspthd/\rtom=0.675^{+0.024}_{-0.021}$ and $0.955\pm0.035$ for the high- and
the low-$\cgal$ subsamples, respectively), even when compared to the splashback
radius corresponding to halos with typical $\Gamma$ for our mass scales. 

\subsection{Does dynamical friction result in a smaller splashback radius?}
\label{sec:dyn}

So far in our analysis, we have identified the splashback radius using the galaxy distribution around
our cluster subsamples. The splashback radius of galaxies could be different
from that of dark matter due to dynamical friction acting on the subhalos that
host our galaxies, provided these subhalos are sufficiently massive \citep{Adhikari:2016}. In
what follows, we show that the steepening of the three dimensional density
profiles for both matter and subhalos that host our fiducial photometric sample
of galaxies are expected to occur at similar locations. 

For this purpose, we again make use of the halo and subhalo catalogs from the
MDPL2 simulation. We match the cumulative abundances of dark matter
subhalos
as a function of ${\vpeak}$ (the maximum circular velocity of a halo throughout
its entire history) and that of our photometric galaxies as a function of their
magnitude, to obtain an estimate of the ${\vpeak}$ of subhalos hosting our
galaxies (see Appendix~\ref{app:amatch}). The subhalos that host our fiducial subsample of
photometric galaxies approximately correspond to subhalos with ${\vpeak}>135
\kms$, while the brighter subsamples correspond to subhalos with $\vpeak >175
\kms$ and ${\vpeak}>280 \kms$, respectively. 

For this analysis, we use the $z=0$ particle snapshot of the
simulation\footnote{Ideally we would have liked to also carry out this exercise
near $z=0.24$, but we had only the $z=0$ particle snapshot available.}. We use all halos
identified by the 6d phase space halo finder ROCKSTAR \citep{Behroozi:2013} in
the $z=0$ snapshot with halo mass, ${\mtom}$, above $8.5\times10^{13}\msunh$ as
our sample of galaxy clusters. We subdivide these in bins of $\Gamma=\Delta\log
\mvir/\Delta\log a$, and compute the three-dimensional density profile of matter
around them. The derivative for $\Gamma$ for this particular snapshot was
computed between $z=0$ and $z=0.5$ \citep{Diemer:2014, More:2015}.
The logarithmic slope of the matter density profile around the cluster samples
are shown in the different panels of Figure~\ref{fig:sh_matter_rsp_3d} using a
black solid line. For reference we also show the expected locations of the
splashback radius for each of the subsamples, using Equation~\ref{eq:meanMDK15}.
The fitting function seems to capture the trend observed for the splashback
radius of dark matter as a function of the accretion rate in the simulation
reasonably well (within 5 percent).

In the same figures we also show the logarithmic slopes of subhalo distributions
around galaxy cluster halos for different ${\vpeak}$ thresholds obtained from
our simple subhalo abundance matching method. We observe that the locations of
the steepest slope for subhalos with the lowest ${\vpeak}$ threshold is similar
to that in dark matter within 5 percent for all the $\Gamma$ bins shown in the
figure. Thus the location
of the splashback radius is not expected to be significantly different for
subhalos hosting our fiducial sample of photometric galaxies. As we consider
${\vpeak}$ thresholds corresponding to our brighter subsamples, we see effects of
dynamical friction acting on the subhalos \citep[see also][]{Jiang:2014}. The
splashback radius of these larger subhalos systematically shifts to smaller
values with increasing $V_{\rm peak}$ threshold. 

We have tried to maximise the effect of dynamical friction in the above exercise
by not considering scatter between the luminosity of galaxies and the $\vpeak $
of their subhalos while performing abundance matching. We do not see a large
shift in the splashback radius of the photometric galaxies around any of our
cluster subsamples as a function of their magnitude threshold. However, this
does not imply that our data rule out dynamical friction acting on the brighter
sample of photometric galaxies. It is quite likely that there is a reasonably
large scatter between the magnitude and ${\vpeak}$, which can wash out the
dynamical friction effect. 

\subsection{Background subtraction and mis-centering uncertainties}
\label{sec:background}

We have used the number density profiles around random points to subtract the
uncorrelated galaxies in the background and the foreground of our subsamples.
We have also tested how residual uncertainties in background subtraction can
affect our results. As an initial test of uncertainties in the background which
are constant with radius, we have added in a constant parameter to our model for
the projected number density profiles. Even after marginalizing over such a
parameter, we obtain values for the splashback radii and its uncertainty which
are very consistent in two dimensions, and virtually identical in three
dimensions. However, there could be additional background uncertainties which
vary with the projected distance.

For example, we expect that the clusters in our subsamples will cause the galaxies in the
background to be magnified \citep[see e.g.,][]{Umetsu:2011}. We explore the
changes to the background due to cluster magnification in
Appendix~\ref{app:mag}, and find that the splashback radius is not affected
even after applying a conservative correction to the background due to the
magnification of the clusters. The sky subtraction around bright or highly
clustered objects can also potentially affect the photometry of galaxies and
hence the background objects in clusters \citep{Aihara:2011}. This can also
partly cancel the magnification effect, as it reduces the number density of
background galaxies in clusters.

Mis-centering of central galaxies in  \redms clusters could affect the profiles
and our estimates of the splashback radius. There are two kinds of
mis-centering: first, a galaxy may be mis-classified as a central by the cluster
finder and second, the central galaxy may be physically displaced from the
potential minimum of the cluster around which all galaxies orbit. To test for
the first kind of mis-centering, we have restricted our model fits to scales
$>400\kpch$ or to using clusters where the most probable central galaxy has
$\pcen>0.9$. These restrictions produce fit parameters consistent with those
listed in Table~\ref{tab:post}, especially the position of the splashback
radius, within the reported uncertainties. 

To test for the second kind of mis-centering, we have also considered all halos
from the MDLP2 simulation used in the previous section, and displaced 40 percent
of these halos in their positions with an offset drawn from a multivariate
Gaussian distribution with standard deviation equal to
$400\kpch$\footnote{This assumes that the \redms centering probabilities are
unreliable and the centering algorithm performs as badly as selecting the
brightest galaxy in the cluster, which could result in 40
percent mis-centering fraction \citep{Skibba:2011}.}. In each panel
of Figure~\ref{fig:sh_matter_rsp_3d}, we have additionally included a dashed
line which shows the slope of the logarithmic density profiles around such a
sample of halos. We find that, as expected, in all cases the splashback radius
would be overestimated by $\gtrsim 20\%$, an effect which goes in the opposite
direction required to explain a smaller splashback radius. Moreover, the change
of slope around the splashback radius is much less pronounced and overall shape
of the profile is significantly modified.  This is in contrast with the good
agreement we find between the {\it shapes} of the predicted and observed slope
profiles.  

\subsection{Could averaging effects result in a smaller splashback radius?}
\label{sec:averaging}

We have used the average halo mass of our subsamples as inferred from weak
lensing to obtain the average $\rtom$ of our cluster halos to compare with the observed values of the
splashback radius, $\rspthd$. Could the difference between the $\rspthd/\rtom$
seen in observations and that predicted from simulations arise due to the finite
width of the halo mass distribution? We considered the distribution of halo
masses resulting from a threshold sample with the same average halo mass as our
cluster subsamples. For such samples, we find that the difference between
$\ave\rtom$ and that inferred from the average halo mass is different by only
$\approx 3$ percent, whereas the discrepancy we observe is much
larger\footnote{Using a halo mass sample with a threshold in $\nsat$ as in our
data, also does not affect this conclusion.}.

We have also verified that the location of the splashback radius for a threshold
mass sample does not result in a smaller inferred splashback radius compared to
the expectation based on using the average halo mass, and the average mass
accretion rate onto the halo samples. These tests confirm that the smaller
value for the splashback radius we observe is not likely to be a result of some
averaging effects.

\begin{figure*} 
\centering{
\includegraphics[scale=0.8]{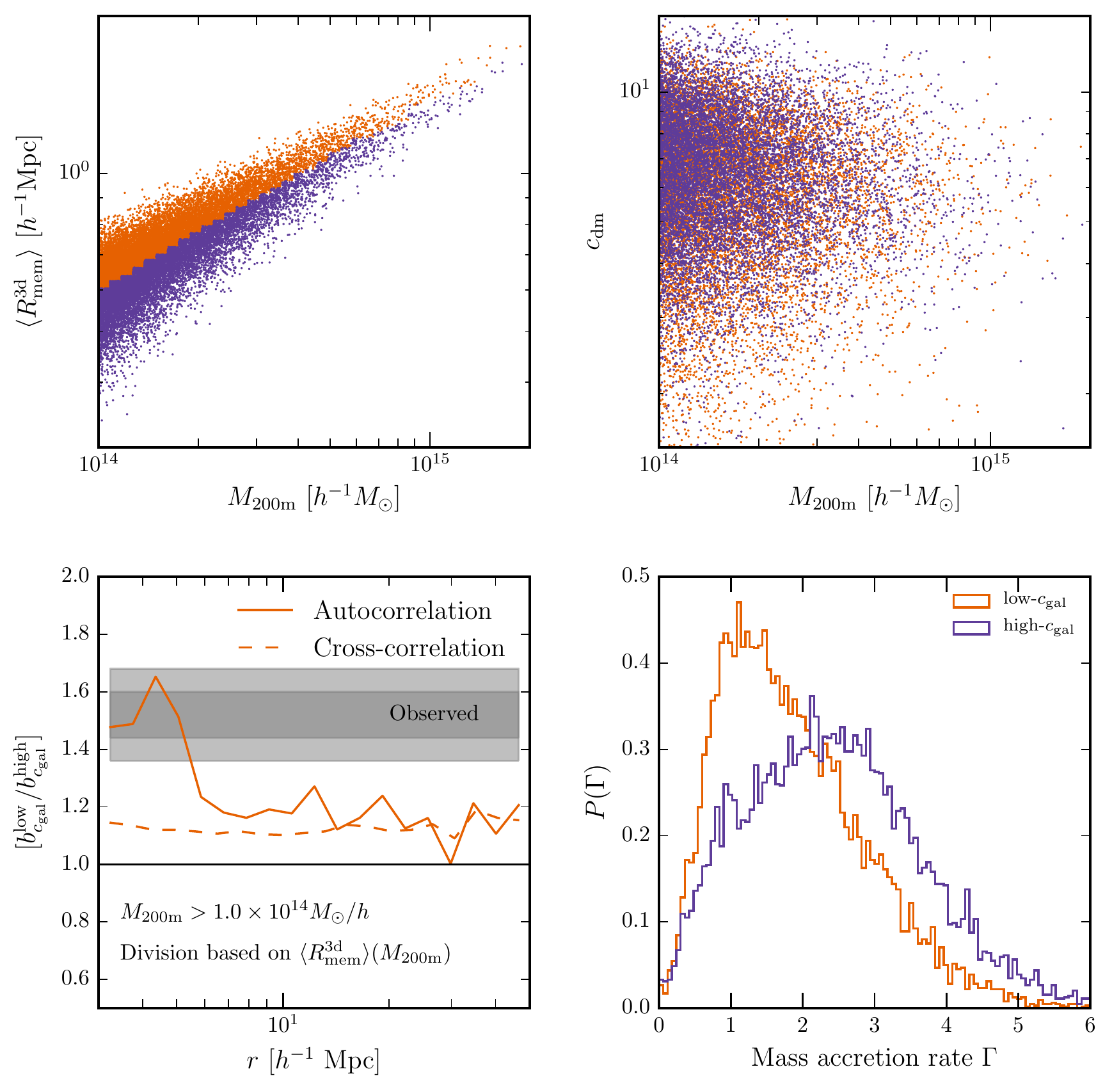}}
\caption{
Expectation for halo assembly bias from simulations. {\it Top left panel}:
The average cluster centric distance of member subhalos in cluster sized halos
in MDPL2 simulation as a function of halo mass. We use the radius to divide the
cluster sample into two at fixed mass. The two subsamples are shown in orange and purple
colors. {\it Top right panel}: The same subsamples as in the top left panel but in the
dark matter concentration-halo mass plane. The subhalo distribution seems to
 have very little correlation with the dark matter distribution.
 {\it Bottom left
panel}: The ratio of the halo biases of subsamples of halos with masses
$10^{14}\msunh$ split using the average distance of their subhalos from
 their centers, $\ave{\rmem^{\rm 3d}}(M_{\rm 200m})$. The bias ratio obtained from
 cluster-cluster auto-correlations is shown with a solid line, while the dashed
 line corresponds to the ratio obtained from the cross-correlations of
 cluster-scale halos with all subhalos with $\vpeak>135\kms$ that are selected
 to mimic the fidual photometric galaxies used in our analysis as in
 Figure~\ref{fig:obs_all}. {\it Bottom right panel}: The distribution of the
 mass accretion rates $\Gamma$ for the two subsamples.
} 
\label{fig:assembly_bias_expect}
\end{figure*}

\subsection{What is the systematic error in the weak lensing halo masses?}

Our conclusion that the observed splashback radius is smaller than the
expectation from simulations is based on the comparison with the virial radius
for weak lensing halo mass inferred by \citet{Miyatake:2016}. However,
\citet{Miyatake:2016} assumed a $\delta$-function distribution in halo masses to
model the weak lensing measurements for each cluster subsample and infer the
average halo mass. Such a simplified fitting to the weak lensing surface mass
density profile is expected to underestimate the virial halo mass by $\sim10\%$
compared to the mean halo mass of clusters in the sample
\citep{Mandelbaum:2005,Becker:2011,Niikura:2015}, or $\sim3\%$ in radius. The
difference between the measured and expected splashback radii is much larger
than such systematic error, and if at all will increase the inconsistency rather
than decrease it. The statistical error in the weak lensing masses is less $10$
percent, so even 2-$\sigma$ deviations will result in only $\sim7$ percent
change in the expectation in the splashback radius. The errors in the weak
lensing masses have also been folded in when computing the error on the observed
$\rsp/\rtom$ shown in Figure~\ref{fig:derivmdk}.

\subsection{Is the halo assembly bias signal consistent with expectations?}

Our observational results indicate that the galaxy cluster subsample with lower
concentration of member galaxies has a larger splashback radius (lower accretion
rate $\Gamma$), and a larger halo bias. Is the sense and the amplitude of the halo
assembly bias signal we see consistent with expectations from cosmological
simulations of cold dark matter?

Various proxies such as formation time scales of halos or their dark matter
concentrations have been used in the literature to quantify halo assembly bias.
The sense of the halo assembly bias effect varies depending upon the assembly
proxy used. For example, \citet{Gao:2005} find that halo assembly bias is
strongest for low mass galaxy scale halos, and that the earliest forming halos
cluster more strongly than the average for their halo masses. However, they find that
the effect almost disappears on the mass scales we consider in our paper. On the
contrary, when the concentration of halos is used as a proxy, halo assembly bias
manifests itself at both galaxy scales and galaxy cluster scales
\citep{Wechsler:2006}. At galaxy scales (masses below $10^{12}\msunh$) high
concentration halos (which form earlier) have a larger halo bias, but the trend
reverses on galaxy cluster scales, as expected from the relation between mass
accretion history and the curvature of initial density peaks \citep{Dalal:2008}.

\citet{Li:2008} explored eight different definitions of halo assembly and found
that the connection between formation time and assembly bias of halos can be
totally washed out or even reversed depending upon the proxy used. Member galaxy
concentration has never been explored previously in the literature as a proxy
for assembly history. Therefore, we use the MDPL2 simulation to explore the
extent of the assembly bias expected when using it as a proxy. We ignore all
observational complications, and investigate how the clustering of halos at
fixed halo mass varies as a function of the concentration of the subhalo
distribution belonging to the halos. 

We use all isolated halos with ${\mtom}>10^{14}\msunh$ at $z=0.248$ from MDPL2.
This threshold in halo mass at $z=0.248$ allows us to match the average weak lensing mass of
our cluster subsamples\footnote{Note that this threshold is slightly larger
than the threshold used used in Section~\ref{sec:dyn} at $z=0.0$. Both
thresholds ensure that we match the average weak lensing mass from observations.}. To compute the average projected separation between
cluster members and halo centers, we use all subhalos with ${\vpeak}>135
\kms$, similar to the threshold used for the faintest of our
photometric galaxies. We divide the sample into halves based on the
3-d cluster centric distance of the subhalos of each cluster as shown in the top
left panel of Figure~\ref{fig:assembly_bias_expect}. The top right panel of the
figure shows the scatter plot of dark matter concentrations of our halos given
their halo masses and implies that the concentration of dark matter and
concentration of subhalos are not well correlated. 

The ratio of the halo bias obtained from the auto-correlation function of
cluster halos in the two subsamples is shown in the bottom left panel of
Figure~\ref{fig:assembly_bias_expect} with the orange solid line, while that
obtained from the cross-correlation with halos with $\vpeak>135\kms$ is shown
with a dashed line. For comparison, the observed ratio between the biases of our
cluster subsamples is shown with the grey shaded region.  Subsample divisions
based on $\ave{{\rmem^{\rm 3d}}}(\nsatthd)$, give very similar amplitude for
this ratio at larger radius when using auto-correlations, but with weaker scale
dependence.  Finally, the bottom right panel shows the distributions of the mass
accretion rates onto these clusters.

The halo assembly bias signal we observe in data is in the same sense, albeit
stronger, than that seen in simulations\footnote{The preliminary
investigations mentioned in \citet{Miyatake:2016} which seemed to suggest larger
assembly bias signal, were mistakenly performed using a larger mass threshold by
SM. The results presented in this paper override those preliminary
investigations.}. The division in the mass accretion rates is also in the same
sense as required to explain the splashback radius measurements seen in the
data. These results from simulations show that it is possible to obtain samples
of halos with a lower average accretion rate that have a larger clustering
signal as seen in our observations.

Once projection effects are accounted for (as in Section~\ref{sec:projection}),
and the sample is divided based upon $\ave{{\rmem^{\rm 2d}}}(\nsattod)$, an even smaller
value for the halo assembly bias signal is obtained in simulations. Given that
the strength of the halo assembly bias signal increases with halo mass, one
could potentially reproduce the result by using larger mass scales in
simulations. However this will further worsen the problem of the smaller
splashback radius. Characterization of the halo assembly bias signal with
different richness threshold samples and theoretical investigations using larger
simulations are currently ongoing and will be reported in a future publication.

\subsection{Projection effects}
\label{sec:projection}

\begin{figure*} 
\centering{
\includegraphics[scale=1.0]{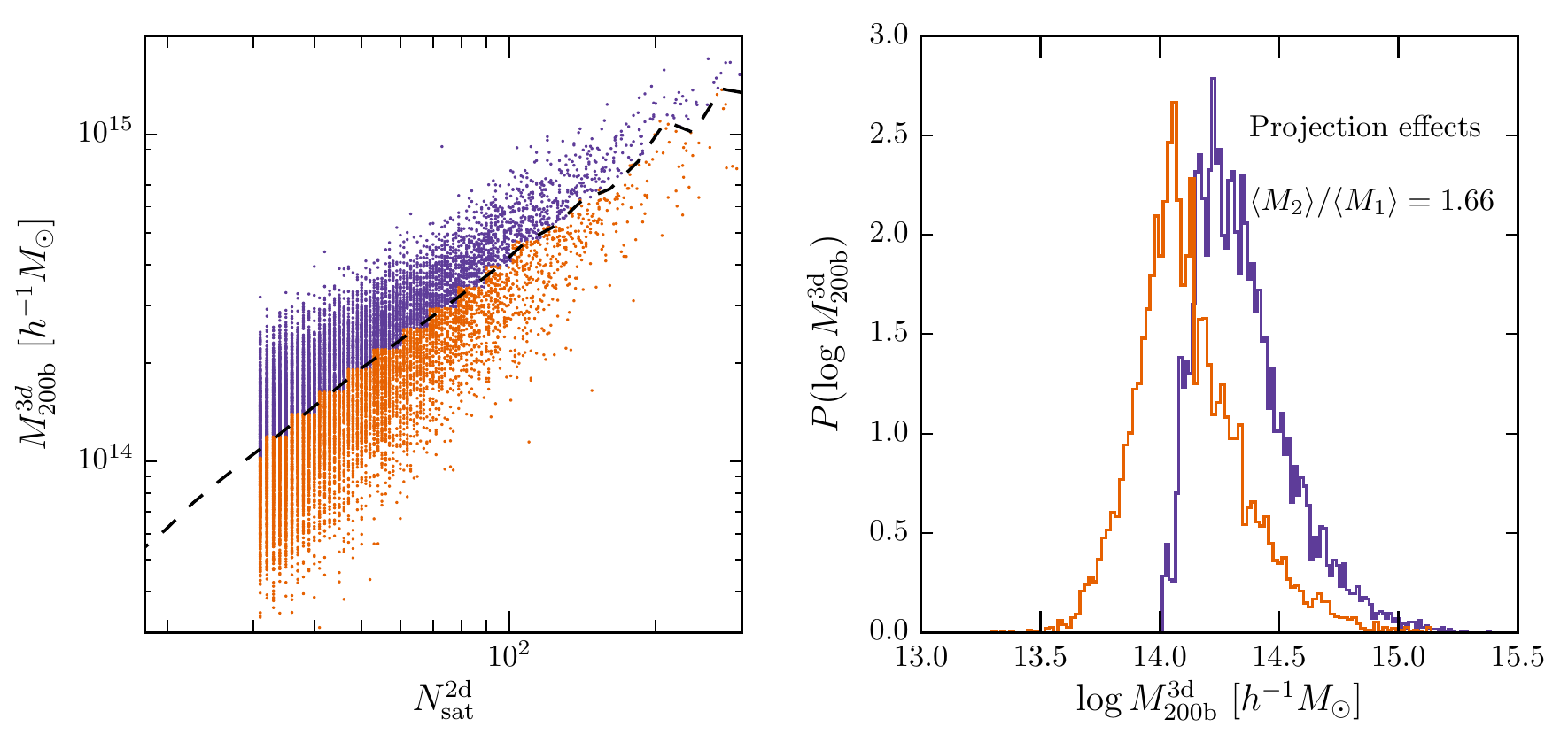}}
\caption{
        Estimates of projection effects. Left panel: The scatter plot of
        $\mtom$ of halos as a function of their richness determined in
        projection. The richness of cluster halos in projection is computed by
        assigning smaller (sub)halos as satellites if they lie within a projected
        radius of $\rtom$ and within $\pm50 \mpch$ of a larger halo along
        the line-of-sight. To maximize projection effects, we assume the two
        subsamples separated by the median ${\mtom}-\nsattod$ relation
        shown by the dashed line. Right panel: The halo mass distributions of
        the two cluster subsamples shown in the left panel. The average halo
        masses are expected to be different by at most a factor of $1.66$ when we
        maximize the projection effects.
}
\label{fig:projection}
\end{figure*}

One possible way to explain the different splashback radii and halo biases for our
two cluster subsamples, could be that the two SDSS cluster subsamples in reality have
different masses. The weak lensing mass estimates, taken at face value, restrict this
possibility. However, the weak lensing signal is only sensitive to the
projected mass distribution. The cluster subsample identification also has a
very poor resolution in the line-of-sight direction. Therefore, we explore the
possibility that our subsamples indeed have different $3$-d masses regardless of
the similarity of the weak lensing signal, and the richness of the clusters.

To address the magnitude of projection effects, we again resort to the MDPL2
simulation. We consider all halos with  ${\mtom}>10^{13} \msunh$ as potential
clusters identified from the photometric data\footnote{We have chosen a slightly lower halo mass threshold for
this test since projection effects can upscatter smaller halos to have a larger
richness than expected for their halo masses.}. We associate all halos/subhalos with ${\vpeak}>135 \kms$ to be
satellite galaxies of these potential galaxy clusters if they lie within a
projected radius of $\rtom$ from their centers, and have a line of sight
separation less than $50 \mpch$. The projected separation lengths of $\pm 50
\mpch$ was obtained by considering the scatter in the colors of red sequence
galaxies used to identify \redms galaxy clusters, and the amount of variation of
these colors as a function of redshift \citep[see Figures 1 and 7
in][]{Rykoff:2014}.  If subhalos can belong to two clusters after the
projection, we assign it to the most massive halo. We do not redefine or
recompute the cluster centers after these projections as we expect that the
\redms centering algorithm is less likely to identify the galaxies in projection
as centrals for the \redms clusters. 

In the left hand panel of Figure~\ref{fig:projection}, we show the scatter plot
of $\mtomthd$ as a function of $\nsattod$, i.e., the
number of satellites associated with these cluster sized halos after the
reassignment described above. To maximize the strength of projection effect, we
make the extreme assumption that our two cluster subsamples select the upper and
the lower half of $\mtomthd$ at fixed $\nsattod$. The
distributions of the $\mtomthd$ selected in this manner are
displayed in the right hand panel of Figure~\ref{fig:projection}. The average
halo masses corresponding to the two distributions are different by $\sim62\%$,
which implies a difference of at most $\sim 20\%$ in the radii.  In
contrast, the observed difference in the radii is twice as large, $\sim40\%$. We
expect the extent of projection effects to be lower, given that the average
cluster centric distances for the satellites are not expected to be perfectly
correlated with $\mtomthd$ at fixed $\nsattod$. Furthermore, due to halo
assembly bias effects in simulations, the sample with larger $\mtomthd$ at fixed
$\nsattod$ turns out to have a smaller clustering signal, opposite of what is
required to simultaneously explain both the different splashback radius and the
sense of the assembly bias signal.

Thus, we have established that the two subsamples have splashback radii and halo
biases that differ from one another by an amount that cannot be explained with
the help of projection effects alone. In Appendix~\ref{app:proj}, we show that
both of our cluster subsamples have very similar distributions of the axis
ratios of the light profiles for the most probable central galaxies, similar
ellipticity distributions for satellite galaxies, and similar line-of-sight
velocity dispersions. The similarity of these distributions is inconsistent with
the trends expected, if the projection effects were behind  the large magnitude
of assembly bias signal or the difference in the splashback radii between the
two subsamples divided by $\ave{R_{\rm mem}}$.

\section{Discussion}
\label{sec:discuss}

In Section~\ref{sec:results}, we showed that the surface number density profiles
of galaxies around \redms clusters exhibit steepening characteristic of the
splashback radius expected to arise in dark matter halos due to the caustic
formed by matter at the first apocenter after accretion.  Moreover, we showed
that subsamples of the clusters of the same mass split on concentration of the
galaxy distribution exhibit different splashback radii and different spatial
bias.

Although comparisons with theoretical expectations
presented in Section~\ref{sec:sim_compare} show that the observed profiles and
trends agree with predictions qualitatively, there are significant quantitative
discrepancies. First, the observed splashback radii are significantly smaller than
expected for halos corresponding to cluster masses of our sample, and second, the difference
in biases of cluster subsamples is also larger than expected. Both discrepancies
are very interesting because the location of the splashback radius is a result of 
simple gravitational dynamics and can be both robustly predicted from simulations 
\citep{Diemer:2014,More:2015} and reproduced even in simple spherical 
collapse models  \citep{Adhikari:2014}. Likewise, halo assembly bias is expected
to reflect the statistics of peaks in the initial density field from which
massive cluster-sized halos form \citep{Dalal:2008}. We will postpone the
discussion of the larger magnitude of halo assembly bias to a future paper where
we will quantify with larger simulations the extent of this discrepancy. We
focus on the splashback radius discrepancy below.

We consider four possible explanations for the discrepancies in the splashback
radius -- (a) the halo mass ${\mtom}$ and hence the radius $\rtom$ inferred from
the data is larger than it really is, and this causes our measurements of
$\rsp/\rtom$ to be biased low, (b) the splashback radius of the clusters for a
given accretion rate measured using galaxies is smaller than that expected from
dark matter simulations, (c) the mass accretion rate onto clusters is much
higher than expected for clusters of this mass, and, finally, (d) that new physics in the
form of self interactions of dark matter particles results in decrease of the 
splashback radius. We discuss each of these possibilities in turn.

The ratio between the splashback radius of the two cluster subsamples is roughly
consistent with the best case expectations from the $\Lambda$CDM simulation
of the Planck cosmology (the ratio between the
splashback radii corresponding to the orange and purple stars in
Figure~\ref{fig:derivmdk}). Let us
assume that our observational proxy results in the best case split, a
very optimistic assumption, and that the weak lensing masses are overestimated.
In that case, the discrepancy seen in the splashback radius for the two cluster
subsamples is $\sim1.4$ in radii, thus a factor of $\sim 2.75$ in halo masses,
strongly disfavored by the small errors on the masses from weak lensing. Under
this scenario, we would have uncovered a critical problem for weak lensing
calibration of observable halo mass relations which are important for precision
cosmological measurements using galaxy clusters. Moreover, even if we assume
that the masses are overestimated for both samples, this would mean that the
predicted assembly bias strength would be even lower. Such arguments thus run in
the opposite sense of what is required to explain the large assembly bias
signal. 

If weak lensing systematics can be ruled out, one could try to explain the discrepancy
if the mass accretion rate onto the clusters is normal, but the splashback
radius is indeed smaller than expected. Dynamical friction on subhalos was an
obvious physical effect which could cause this. However, in
Section~\ref{sec:dyn}, we have shown that the subhalos corresponding to the
faintest magnitude bin we consider are not expected to experience a large amount
of dynamical friction. Also, puzzlingly, in the data we do not detect a significant difference in the
splashback radius with the brightness of the photometric sample used. Furthermore, future deeper surveys should be able to rule
out this possibility by using even fainter galaxies. Other astrophysical
possibilities include subtle biases in the photometric galaxy samples, which
lead to a preferential selection of slow moving galaxies in our sample. For
example, ram pressure stripping is expected to be more effective in stripping out gas
and shutting down the star formation in fast moving galaxies, thus removing them
preferentially from our photometric sample. Detection of the splashback radius
of matter in clusters using the weak lensing signal would be able to test this
possibility.

Dark matter self-interactions have long been proposed to alleviate problems on
small scales in the standard cosmological model \citep[see
e.g.,][]{Spergel:2000}. Under certain conditions discussed below the drag force
due to interactions between dark matter particles of subhalos and cluster halo
could lead to loss of orbital energy by subhalos even on the first crossing,
thereby reducing the splashback radius. 

For isotropic elastic scattering, we do not expect dark matter self-interactions
to significantly affect the splashback feature, because the upper limits on such
an interaction cross-section are sufficiently stringent to ensure that most dark
matter particles do not experience any scattering events during a single orbit
\citep{Gnedin:2001, Randall:2008}.  Of the few subhalo particles that do
scatter, most are ejected from their subhalos, since the orbital velocities of
subhalos within massive hosts are typically larger than the internal escape
velocities of those subhalos. Therefore we would expect evaporation of subhalo
masses, without a significant drag for isotropic scattering.

On the other hand, if dark matter self-interactions are anisotropic, with large
cross-sections for small angle scattering and low cross-sections otherwise, then
the momentum transfer during dark matter interactions
may not necessarily be large enough to ensure ejection. The small angle
scattering cross-sections could then be large enough for dark matter particles
to experience frequent interactions and yet obey the bounds on subhalo
evaporation. The subhalos would experience a net deceleration given by
\begin{equation}
        d = \frac{\rho(r, t) v(t)^2 \sigma_{\rm tr}}{2\mdm}
\end{equation}
where $v(t)$ is the relative velocity of the subhalo, $\rho(r, t)$ is the time
dependent density profile
of the cluster halo, $\mdm$ is the mass of the dark matter
particle and $\sigma_{\rm tr}$ is the momentum transfer cross-section
\citep{Kahlhoefer:2014}.

We have carried out simple analytical calculations based on a spherical collapse
model similar to \citet[][see also \citealt{Adhikari:2016}]{Adhikari:2014}, but
including a velocity-dependent drag term of the above form. We find that the
momentum transfer cross-section required to reduce the splashback radius  by
$\approx 20\%$ can range from $1-20\cmsqpg$ depending upon the pericenter of
accreting halos on their first passage through the halo \citep{More:2016b}. The
ambient dark matter density, the relative velocity, hence the resultant drag,
reach a maximum at the pericenter. Therefore, a proper treatment of the orbital
parameters of subhalos expected in the standard structure formation model is
required to determine the effects of dark matter self-interactions on the
splashback radius \citep{Jiang:2014b}. We defer such investigations to a future
paper.

Although the existing constraints on such scenarios are pretty weak, recent
discovery of galaxy displacement with respect to its subhalo in one of the
clusters \citep[][]{Harvey:2015} could be a signature of self interaction
\citep[with cross section consistent with that required to explain $R_{\rm sp}$
discrepancy, see][]{Kahlhoefer:2015}.  Numerical simulations of this type of
dark matter
self-interaction, similar to the simulations performed for hard-sphere
interactions \citep{Elbert:2015}, would be required to refine the estimate of
the cross-sections stated above further.

Note that even if the self interactions will ultimately not turn out to be the
explanation for the splashback radius discrepancy, our analysis shows that
precise measurements of galaxy distribution in clusters could provide valuable
and competitive constraints on the cross section of dark matter self
interaction. 

If we assume that the differences in the splashback radius we find are not due to
the above possibilities and we trust simple dynamics within the gravitational
potential
 of halos, then our measurements of the smaller splashback radius would
either require a different phase space structure in the outskirts of cluster
halos or extreme mass accretion rates onto our cluster subsamples. The former
possibility requires the velocities of material in the infall streams to be
about 25-30 percent smaller. For the latter possibility, our parent sample of
clusters requires values of $\Gamma\sim4$, while the high-$\cgal$ sample prefers
values of $\Gamma\sim4$. This would represent a serious challenge to the
standard cosmological model. Whether modifications to gravity could achieve such
values, while still obeying the bounds from galaxy cluster abundances, needs
further careful evaluation.

\section{Summary}
\label{sec:summary}

We have used SDSS \redms galaxy clusters and photometric galaxies around them
to observationally investigate the boundaries of galaxy clusters, their relation
to assembly history, and to halo assembly bias on galaxy cluster scales. For
this purpose, we have considered two cluster subsamples defined in
\citet{Miyatake:2016} which share identical richness and redshift distribution,
but have different internal distributions of cluster members. These subsamples
were shown to have similar halo masses as inferred from weak gravitational
lensing, but have different large scale bias as measured from their large scale
weak lensing and auto-correlation signals \citep{Miyatake:2016}. Our
results can be summarized as follows:

\begin{enumerate}

\item We detect the surface number density profiles of SDSS photometric galaxies
        with $^{0.1}M_i-5\log h<-19.43$ around both our cluster subsamples. The
        surface densities show a sharp steepening around scales of $1 \mpch$.

\item We modeled the two surface density profiles using the profile advocated by
        \citet{Diemer:2014}, to infer the location of the steepening in
        projected and real space for both the cluster subsamples. The steepening
        of the surface density profiles occurs at significantly different
        locations for the two cluster subsamples. We interpret the steepening as
        the location of the splashback radius for these galaxy clusters. We
        attribute the difference in the splashback radii for the cluster
        subsamples to arise as a result of different accretion rates onto the
        cluster subsamples. This implies a different assembly history for the two
        cluster subsamples.
        
\item Using simple subhalo abundance matching, we investigated whether dynamical
        friction affects the location of the splashback radius of subhalos
        expected to host our galaxies compared to that of dark matter. For the
        fiducial subsample, we found that the two radii should coincide within
        5 percent. Observationally, the location of the steepening of the
        galaxy density profiles for the two cluster subsamples does not change
        significantly even for photometric galaxies one or two magnitudes
        brighter than our fiducial sample.

\item We showed that the amplitude of clustering of photometric galaxies around
        our two subsamples of galaxy clusters are different by $6.6\sigma$. We
        have thus detected halo assembly bias, a difference in the clustering
        amplitude of cluster scale halos with the same mass and different
        assembly histories.

\item We showed that projection effects could at most cause the two subsamples
        to have masses different by a factor of $1.6$, which is smaller than
        the difference required to explain the difference in the splashback
        radii of the two subsamples, or the different halo biases.

\item Using a large cosmological $\Lambda$CDM simulation, we have shown
        qualitative agreement between the trends in the splashback radius and
        halo assembly bias, as seen in observations. However, the splashback
        radii of the two cluster subsamples seem to be smaller, while the
        assembly bias effect larger, than naive expectations. The tests
        presented in the paper show that none of the systematics are large
        enough alone to resolve the discrepancies. If astrophysical systematics
        related to weak lensing, optical cluster finding, and projection effects
        can be conclusively ruled out, it will imply either a discrepancy
        between the observed accretion rates onto clusters from the expected
        ones, or may hint to a possible interactions in the dark matter sector,
        both remarkably interesting possibilities. 

\end{enumerate}

\section{Acknowledgments}

Funding for the SDSS and SDSS-II has been provided by the Alfred P. Sloan
Foundation, the Participating Institutions, the National Science Foundation, the
U.S. Department of Energy, the National Aeronautics and Space Administration,
the Japanese Monbukagakusho, the Max Planck Society, and the Higher Education
Funding Council for England. The SDSS Web Site is http://www.sdss.org/.

The SDSS is managed by the Astrophysical Research Consortium for the
Participating Institutions. The Participating Institutions are the American
Museum of Natural History, Astrophysical Institute Potsdam, University of Basel,
University of Cambridge, Case Western Reserve University, University of Chicago,
Drexel University, Fermilab, the Institute for Advanced Study, the Japan
Participation Group, Johns Hopkins University, the Joint Institute for Nuclear
Astrophysics, the Kavli Institute for Particle Astrophysics and Cosmology, the
Korean Scientist Group, the Chinese Academy of Sciences (LAMOST), Los Alamos
National Laboratory, the Max-Planck-Institute for Astronomy (MPIA), the
Max-Planck-Institute for Astrophysics (MPA), New Mexico State University, Ohio
State University, University of Pittsburgh, University of Portsmouth, Princeton
University, the United States Naval Observatory, and the University of
Washington.

The CosmoSim database used in this paper is a service by the Leibniz-Institute
for Astrophysics Potsdam (AIP). The MultiDark database was developed in
cooperation with the Spanish MultiDark Consolider Project CSD2009-00064. The
MultiDark-Planck II (MDPL2) simulation has been performed in the Supermuc
supercomputer at LRZ using time granted by PRACE.

We acknowledge useful discussions with Simon White, Frank van den Bosch, Andrew
Hearin, Andrew Zentner, Erik Tollerud, Shigeki Matsumoto, Justin Khoury, Mark
Trodden, Bhuvnesh Jain, Daisuke Nagai and Uros Seljak.
MT and SM are supported by World Premier International Research Center
Initiative (WPI Initiative), MEXT, Japan, and by the FIRST program
`Subaru Measurements of Images and Redshifts (SuMIRe)', CSTP, Japan. SM,
MT and MO are also supported by Grant-in-Aid for Scientific Research from
the JSPS Promotion of Science (No. 15K17600, 23340061, 26610058 and 26800093),
MEXT Grant-in-Aid for Scientific Research on Innovative Areas (No.
15H05893, 15K21733, 15H05892) and by JSPS Program for Advancing Strategic
International Networks to Accelerate the Circulation of Talented
Researchers. HM is supported in part by Japan Society for the
Promotion of Science (JSPS) Research Fellowships for Young Scientists
and by the Jet Propulsion Laboratory, California Institute of
Technology, under a contract with the National Aeronautics and Space
Administration. AK was supported by  by the Kavli Institute for Cosmological
Physics at the University of Chicago through grant PHY-1125897 and an endowment
from the Kavli Foundation and its founder Fred Kavli. RyM acknowledges financial
support from the University of Tokyo-Princeton strategic partnership grant. RM
acknowledges the support of the Department of Energy Early Career Award program.


\appendix

\section{Testing methods using simulations}
\label{app:fittest}

\begin{figure*} 
\centering{
\includegraphics[scale=1.2]{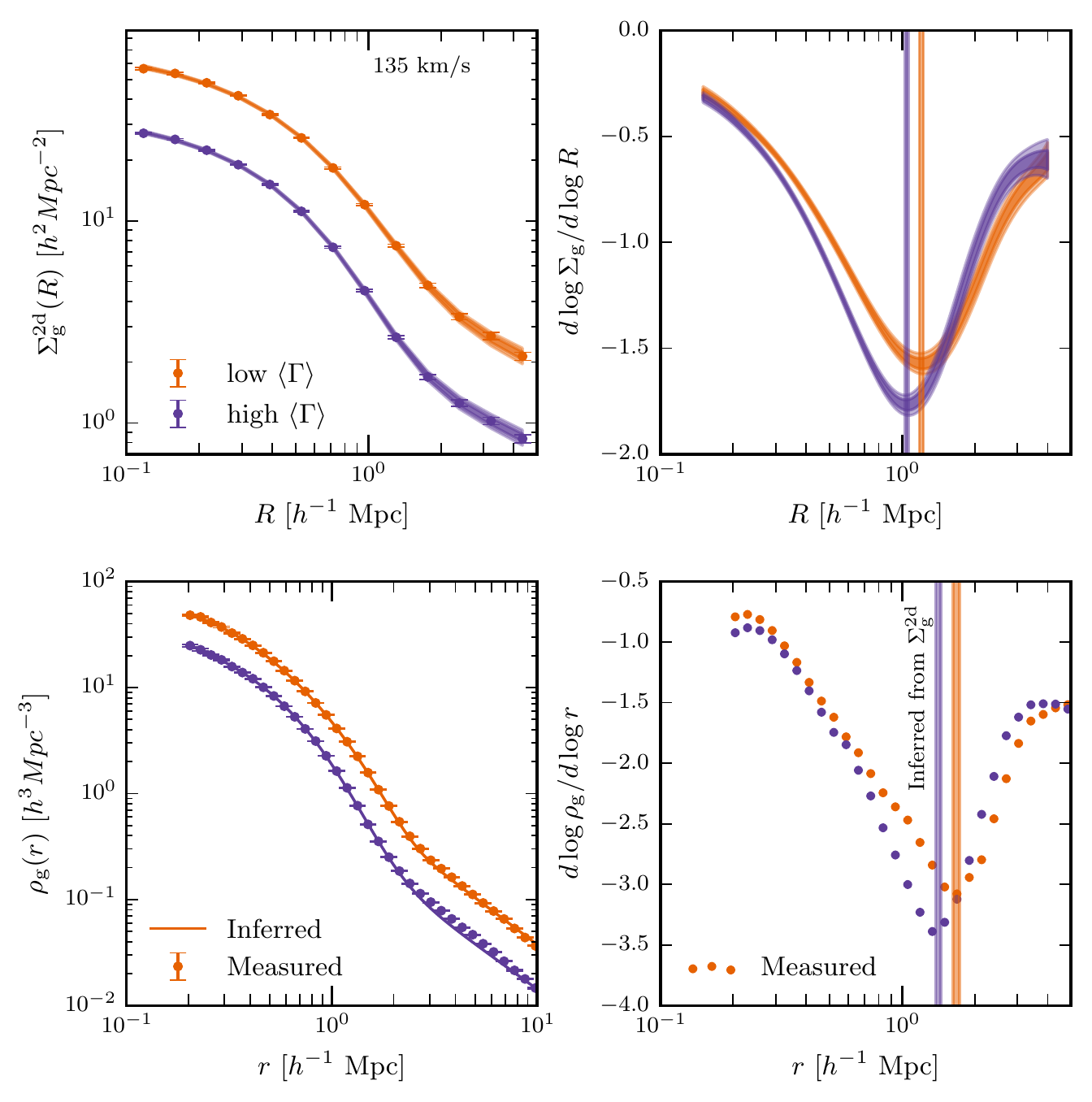}}
\caption{
        Tests of our methodology to determine the splashback radius of clusters.
        Top left panel: The subhalo surface density profiles around high and low
        accretion rate clusters in the MDPL2 simulation at z=0 are shown using
        purple and orange color symbols with errorbars, respectively. The data
        points for the low accretion rate clusters are shifted by a
        factor of 2.5
        for clarity. The 68 and 95 percent confidence intervals based on fitting
        this data using the DK14 model are shown using shaded regions. Top right
        panel: The logarithmic derivative of the surface density profiles around
        the mock cluster subsamples as inferred from a model fit to the data.
        The vertical shaded bands correspond to the location of the steepening
        in two dimensions.
        Bottom left panel: The comparison between the three dimensional subhalo
        density profiles around our mock clusters, as directly measured and as
        inferred from the fits to the surface density profiles from the top left
        panel. Bottom right panel: Comparison between the logarithmic derivative
        of the three dimensional subhalo density profiles around our mock
        clusters, as directly measured and as inferred from the fits to the
        surface density profiles from the top left panel. The 68 percent
        constraints on the location of the splashback radius are shown using the
        shaded vertical bands. Our analytical methods can recover the location
        of the splashback radius in three dimensions.
}
\label{fig:fittest}
\end{figure*}

In this appendix, we use the MultiDark Planck II simulation to test our analysis
methods. In particular, we show that we can recover the splashback radius for
subhalos in three dimensions using the surface density profiles of subhalos
around cluster scale halos. 

We use all halos or subhalos in the halo catalogs with $V_{\rm peak}>135\kms$ at
$z=0$\footnote{See Appendix B for our justification to choose a threshold of
$135\kms$.}. We compute the projected surface density of these subhalos around
halos with masses $M>8.5\times10^{13}\msunh$, where the threshold was chosen to
obtain a sample of halos which share the average weak lensing mass of the \redms
clusters we use in this paper \footnote{This threshold halo mass is similar to
the one used in Section~\ref{sec:dyn} where we tested effects of dynamical
friction using the $z=0$ snapshot.}. We subdivide the sample into two subsamples
with different mass accretion rates, $\Gamma$, using the median mass accretion
rate-halo mass relation $\Gamma(M)$.

We compute the galaxy surface densities around the resultant subsamples by
projecting along the entire $z$-axis in the simulation. We subtract the
background surface density mimicking the procedure used to compute the surface
densities in the data. The errorbars on these surface densities are obtained
using 125 jackknife regions in the simulation box. The resulting surface density
profiles and their errorbars are shown in the top left panel of
Figure~\ref{fig:fittest}. The measurements in the simulations are carried out
using the same binning scheme as employed for the actual data analysis. 

We fit these measured surface density profiles using the projection of the 3d
DK14 density profile with the exact same priors as detailed in
Section~\ref{sec:data}. The 68 and 95 percent confidence intervals on the fit
are shown using the shaded regions in the top left panel of
Figure~\ref{fig:fittest}. The inferred slope of the surface density profiles and
the location of the steepest slope in two dimensions are shown in the top right
panel.

The points in the bottom left panel correspond to the three dimensional density
profiles of these subhalos as measured from the simulation directly. The solid
lines are used to indicate the three dimensional subhalo density profiles as
predicted by the best fit to the subhalo surface density profiles in the upper
panels. Finally, in the bottom right panel we compare the inferred and measured
logarithmic slopes of the three dimensional density profiles. This test shows
that our analysis methods can reproduce the location of the splashback radius by
fitting the two dimensional density profiles of subhalos in the simulations.

\section{Abundance matching constraints on the subhalos hosting photometric
galaxies}
\label{app:amatch}

\begin{figure*} 
\centering{
\includegraphics[scale=1.2]{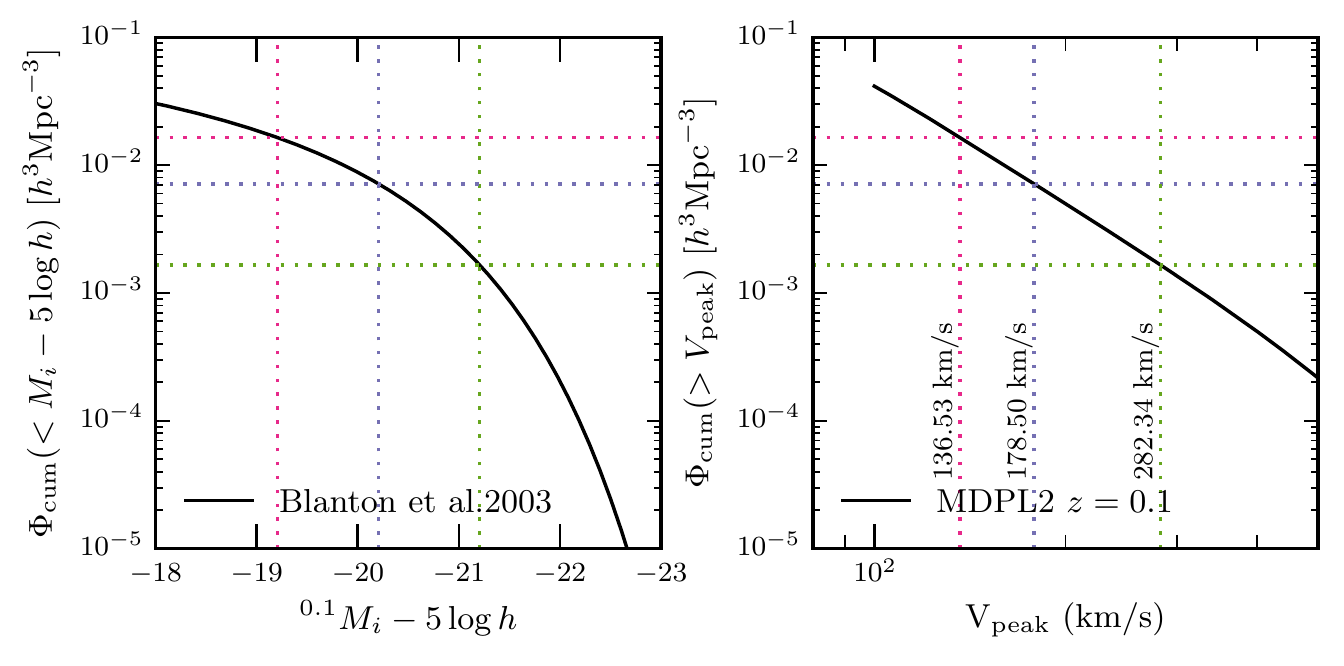}}
\caption{
        Left panel: The cumulative abundance of SDSS galaxies as a function of
        their $i$-band magnitude from \citet{Blanton:2003}. The dotted vertical
        lines correspond to the magnitude limits of our photometric samples of
        galaxies k-corrected and evolution corrected to $z=0.24$, the median
        redshift of the sample. The horizontal lines give the cumulative
        abundances. Right panel: The cumulative abundances of subhalos in MDPL2
        simulation as a function of ${\vpeak}$. The horizontal dotted lines are
        the cumulative abundances of our photometric samples taken from the left
        panel, while the vertical dashed lines give the ${\vpeak}$ thresholds
        corresponding to our photometric subsamples. These $\vpeak$ thresholds
        are used to test dynamical friction effects on the location of the
        splashback radius (see Figure~\ref{fig:sh_matter_rsp_3d}).
}
\label{fig:amatch}
\end{figure*}

We use the surface number density distribution of photometric galaxies around
our cluster subsamples to detect the splashback radius. In this appendix, we use
a simple subhalo abundance matching technique to determine the properties of
subhalos hosting the photometric galaxies we use in this paper. The left hand
panel of Figure~\ref{fig:amatch}, shows  the cumulative abundance of galaxies,
based on the Schechter function fit to the $^{0.1}i$-band luminosity function
for SDSS galaxies obtained by \citet{Blanton:2003} \footnote{The notation
$^{0.1}i$ stands for magnitude in the $i$-band k-corrected to $z=0.1$.}.

We k-correct the absolute magnitude limits of the
photometric galaxies we use, as well as correct them for the luminosity evolution
of galaxies (e-correction) from $z=0.24$, the median redshift of our cluster
subsamples, to $z=0.1$. We approximate the k-correction as
\begin{equation}
        ^{0.1}k(z) = -2.5 \log_{10}\left[\frac{z+1.3}{1.1(0.1+1.3)}\right]\,,
\end{equation}
found by fitting the k-correction as a function of redshift using the SDSS main
sample of spectroscopic galaxies. We have ignored the residual color dependent
scatter in this relation. These $k+e$ corrected magnitude limits for our
photometric subsamples are shown with vertical dashed lines, while the
horizontal dashed lines show their cumulative abundances. We do not assume any
scatter between ${\vpeak}$ and magnitude, to obtain a limit on the maximum
effect that dynamical friction can have, and match these abundances directly to
those of subhalos as a function of ${\vpeak}$. The result of this simple subhalo
abundance matching exercise is shown in the right hand panel of
Figure~\ref{fig:amatch}. 

The abundance matching implied that the subhalos hosting our fiducial subsample
of photometric galaxies approximately correspond to subhalos with ${\vpeak}>135
\kms$, while the brighter subsamples correspond to subhalos with $\vpeak >175
\kms$ and ${\vpeak}>280 \kms$, respectively. In Section~\ref{sec:dyn}, these
abundance matching constraints are used to explore how dynamical friction is
expected to affect the location of the splashback radius for our subsamples.

\begin{figure*} 
\includegraphics[scale=0.85]{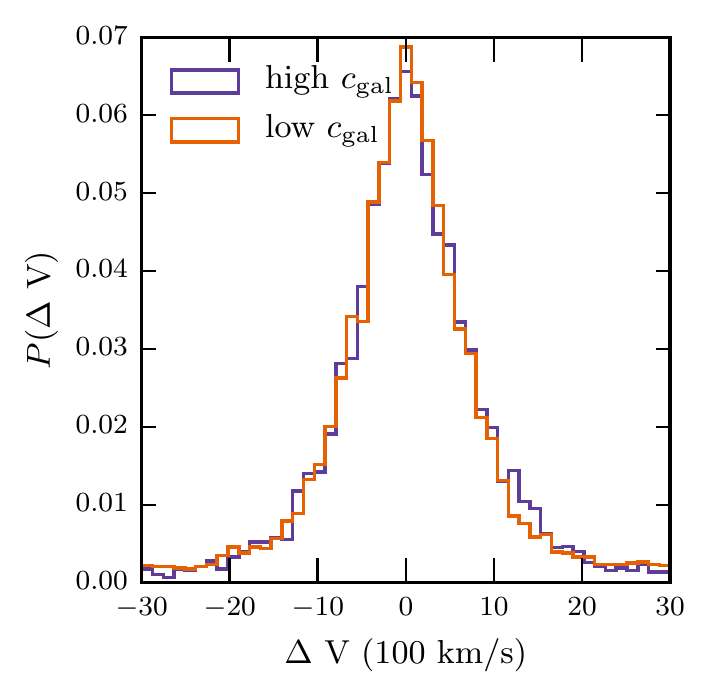}
\includegraphics[scale=0.85]{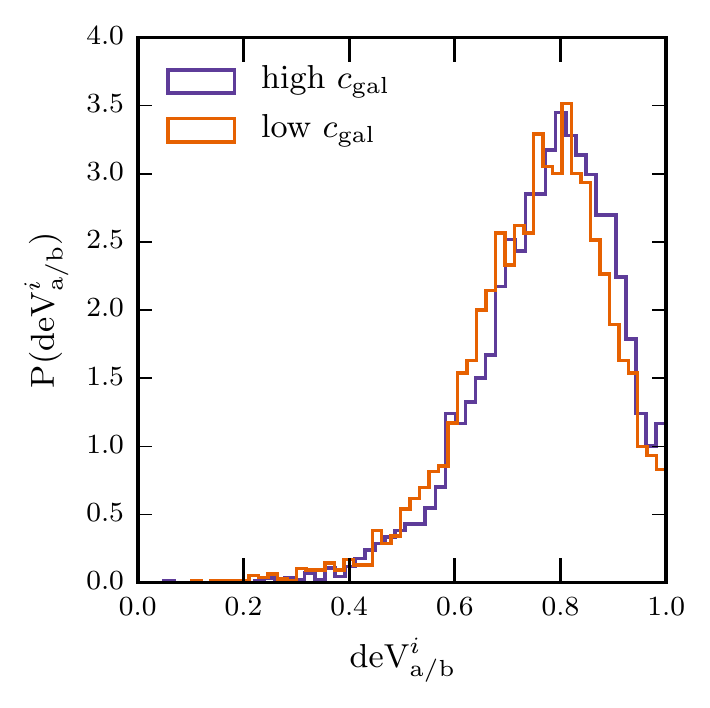}
\includegraphics[scale=0.85]{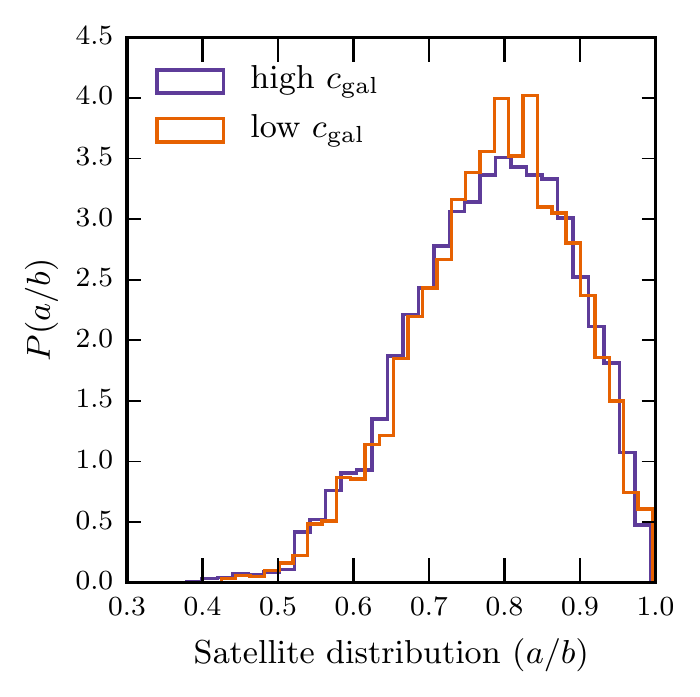}
\caption{
        Left panel: The distributions of line-of-sight velocities of
        spectroscopic galaxies from SDSS around most probable \redms central
        galaxies in the two subsamples.  Middle panel: The distributions of the
        axis ratios of the light profiles in the $i$ band for the most probable
        centrals in the two cluster subsamples are shown. Right panel: The
        distributions of the axis ratios of the satellite distributions around
        the most probable centrals in the two cluster subsamples are shown.  We
        do not see a large differences between our two cluster subsamples for
        any of these statistics.
}
\label{fig:further_tests}
\end{figure*}

\section{Additional investigations regarding projection and halo orientation effects}
\label{app:proj}

Galaxy cluster halos are expected to be triaxial, and if the major axis of
clusters in one of our subsamples were preferentially aligned along the
line-of-sight, we could overestimate the halo mass of that particular subsample
\citep[see e.g.,][]{Oguri:2005}. We have performed the following tests to
evaluate such possibility. Any attempts to explain the halo assembly bias signal
based on projection or orientation effects, should also satisfy these
observational constraints.

We have utilized the spectroscopic coverage of the SDSS-DR12 LOWZ galaxy sample to
investigate the velocity distributions of such galaxies around our cluster
subsamples. If one of our subsamples is heavily affected by projection effects,
then the velocity distributions around the two subsamples should reveal
differences. The \redms cluster catalog contains spectroscopic redshifts for 3037
(2830) most probable central galaxies in the high- (low-) $\cgal$ subsamples. We
show the $P(\Delta V)$ distribution of all spectroscopic galaxies within a
projected radius of $\rtom$ of the cluster center in the left hand panel of
Figure~\ref{fig:further_tests}. We find very little evidence, if any, of a
difference between the two distributions.

The photometric major axis of central galaxies has been reported to have a preferential
alignment with the major axis of the dark matter halo, albeit with a reasonable
scatter \citep[see e.g.,][]{EvansBridle:2009,Ogurietal:2010,Oguri:2012}. If one of our subsamples (the high-$\cgal$ subsample) had
major axis which were preferentially aligned with the line-of-sight, we expect
the BCG ellipticities of that sample to be rounder than the other. In the middle
hand panel of
Figure~\ref{fig:further_tests}, we show the axis ratio distributions of the light distribution
in various SDSS bands as reported by the SDSS photometric pipeline, of the most
probable central galaxies in our two subsamples. These axis ratios were computed
by fitting a deVaucouleurs profile to the two-dimensional image of each object
in each band \citep{Stoughton:2002}. We do not see any strong evidence for a
difference in the axis ratios of the most probable central galaxies between the
two cluster subsamples.

The satellite galaxy distribution is also expected to have a preferential
alignment with the major axis of the dark matter halo \citep[see
e.g.,][]{Zentner:2005, Kang:2007}. We compute the second moment of the projected
satellite galaxy distribution around our cluster subsamples using the membership
probabilities of all member galaxies as reported in the \redms catalog. We use
these second moments to compute the major and minor axis, and their axis ratios,
around the most probable central galaxies.  In the right hand panel of
Figure~\ref{fig:further_tests},
we show the axis ratio distributions of the member galaxy distributions. Again
we see very little difference between the two cluster subsamples.

\section{Magnification due to clusters}
\label{app:mag}

The background density of photometric galaxies behind clusters is expected to be
different from the background density around random points due to magnification
effects by galaxy clusters. We consider a simple estimate of this effect based
on the weak lensing mass estimate of our clusters. The magnification changes the
number counts in two ways: firstly, fainter galaxies can now enter our sample
after being magnified by the cluster, secondly, the background galaxies occupy a
smaller solid angle behind the cluster than in the absence of magnification.

For simplicity, we assume that all our clusters are located at the median
redshift of our subsample. We assume that background galaxies follow the
luminosity function presented by \citet{Blanton:2003}. The number of galaxies
that will be observed at a projected distance $R$ from the cluster is then given by
\begin{equation}
        N(R| z_\rml, z_\rms) = \int dz_\rms \frac{dV}{dz_\rms}
        \frac{1}{\mu(R|z_\rml, z_\rms)} 
        \Phi\left(M<M_{\rm max}'|z_\rms\right)
        \label{eq:mag}
\end{equation}
where $\mu$ denotes the magnification due to the cluster at a projected distance
$R$, and 
\begin{equation}
        M_{\rm max}' = M_{\rm max} - 5.0 \log_{10} (D_l[z_\rms]/
        D_l[z_\rml]) + 2.5 \log_{10} \mu(R|z_\rml, z_\rms)\,.
\end{equation}
Here $M_{\rm max}$ corresponds to a maximum absolute magnitude for galaxies that
end up in our subsample. In the paper, we have considered three different thresholds,
$M_{\rm max}-5\log h=-19.43$, $-20.43$ and $-21.43$. The above equation accounts
for the fact that we have assumed the redshift of the photometric galaxies to be
equal to that of the lens redshift while converting their apparent magnitudes to
absolute magnitudes. We additionally k+e correct $M_{\rm max}$ to $z=0.1$ as the
luminosity function we use is based on such k+e corrected magnitudes. 

\begin{figure} 
\centering{
\includegraphics[scale=1.2]{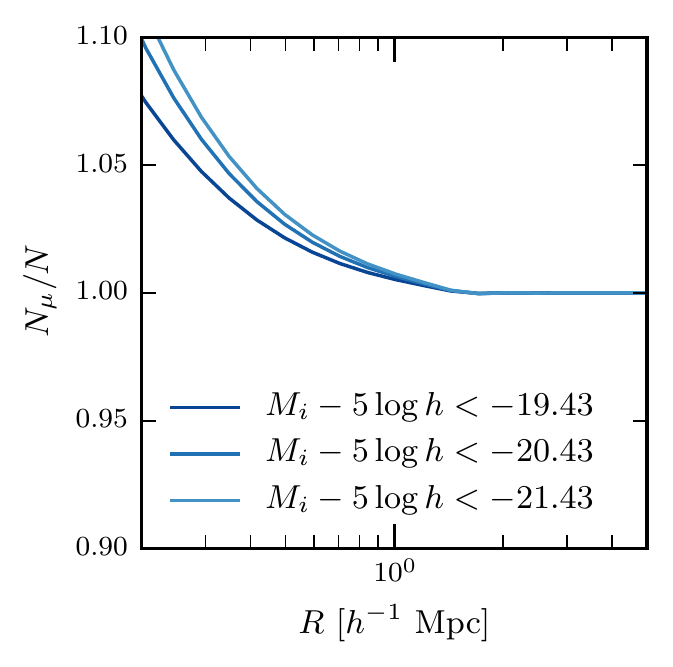}}
\caption{
        Magnification effects on the background estimation: The ratio of the
        number counts of background galaxies behind a cluster compared to that
        around random points, for the three different photometric samples we use
        in the paper. On scales of interest for the splashback radius the
        corrections to the background estimate are at the percent level. We have
        checked that the splashback radius estimates are not affected by such
        small changes to the background.
}
\label{fig:magbias}
\end{figure}

In Figure~\ref{fig:magbias}, we plot the ratio between the number counts
obtained by using Eq~\ref{eq:mag} with $\mu$ obtained for a halo of mass $\mtom$
that matches the weak lensing estimate for our cluster subsamples. We see that
due to the magnification of our clusters, we should be underestimating our
background at smaller radii by 5 to 10 percent, depending upon the photometric
sample under consideration, although the corrections to the background near the
splashback radii is about 1 percent. To check for the systematic effect arising from
magnification effects, we multiply the number counts around random points by this
ratio before subtracting them from the number of galaxies around our cluster
subsamples. Note that this is a conservative estimate of the effect, because
this factor should only be applied to the background galaxies, where as the
number of galaxies around random points also consist of foreground galaxies. We
have confirmed that the location of the splashback radius is insensitive to such
a change. If at all it is expected to be lowered by such a change, given that we
are underestimating the background.

\bibliographystyle{apj}
\bibliography{sf}

\end{document}

%% file: commands.tex
\newcommand{\kpch}{\>{h^{-1}{\rm kpc}}}
\newcommand{\mpch}{\>h^{-1}{\rm {Mpc}}}

\newcommand{\msunh}{\>h^{-1} M_\odot}

\newcommand{\kms}{\>{\rm km}\,{\rm s}^{-1}}

\def\mvir{M_{\rm vir}}

\def\gcm3{\mathrm{g} / \mathrm{cm}^3}
\def\m200m{M_{\rm 200m}}

\def\gtsima{$\; \buildrel > \over \sim \;$}
\def\ltsima{$\; \buildrel < \over \sim \;$}
\def\prosima{$\; \buildrel \propto \over \sim \;$}
\def\gsim{\lower.7ex\hbox{\gtsima}}
\def\lsim{\lower.7ex\hbox{\ltsima}}
\def\simgt{\lower.7ex\hbox{\gtsima}}
\def\simlt{\lower.7ex\hbox{\ltsima}}
\def\simpr{\lower.7ex\hbox{\prosima}}

\def\rmg{{\rm g}}

\def\rml{{\rm l}}

\def\rmo{{\rm o}}

\def\rms{{\rm s}}

%% file: macros.tex
\def\rs{r_{\rm s}}
\def\rhos{\rho_{\rm s}}
\def\se{s_{\rm e}}
\def\ftrans{f_{\rm trans}}

\def\rsp{R_{\rm sp}}

\def\cgal{c_{\rm gal}}

\def\omm{\Omega_{\rm m}}
\def\pcen{p_{\rm cen}}
\def\pmem{p_{\rm mem}}
\def\sigmag{\Sigma_{\rm g}}
\def\rhog{\rho_{\rm g}}
\def\rhoginner{\rho_{\rm g}^{\rm inner}}
\def\rhogouter{\rho_{\rm g}^{\rm outer}}
\def\rt{r_{\rm t}}
\def\rout{r_{\rm out}}
\def\mdm{m_{\rm dm}}

\def\nsat{N_{\rm sat}}
\def\nsattod{N_{\rm sat}^{\rm 2d}}
\def\nsatthd{N_{\rm sat}^{\rm 3d}}
\def\rmem{R_{\rm mem}}

\def\rsptod{R_{\rm sp}^{\rm 2d}}
\def\rspthd{R_{\rm sp}^{\rm 3d}}
\def\mtom{M_{\rm 200m}}
\def\mtomthd{M_{\rm 200m}^{\rm 3d}}
\def\rtom{R_{\rm 200m}}

\def\mvir{M_{\rm vir}}

\def\vpeak{V_{\rm peak}}
\def\cmsqpg{{~\rm cm^2g^{-1}}}

\newcommand{\redm}{redMaPPer}
\newcommand{\redms}{redMaPPer }

\def\ave#1{\left\langle #1 \right\rangle}

%% file: citation_fix.tex
\usepackage{etoolbox}
\makeatletter
\patchcmd{\NAT@citex}
  {\@citea\NAT@hyper@{\NAT@nmfmt{\NAT@nm}\NAT@date}}
  {\@citea\NAT@nmfmt{\NAT@nm}\NAT@hyper@{\NAT@date}}
  {}
  {}
\patchcmd{\NAT@citex}
  {\@citea\NAT@hyper@{%
     \NAT@nmfmt{\NAT@nm}%
     \hyper@natlinkbreak{\NAT@aysep\NAT@spacechar}{\@citeb\@extra@b@citeb}%
     \NAT@date}}
  {\@citea\NAT@nmfmt{\NAT@nm}%
   \NAT@aysep\NAT@spacechar%
   \NAT@hyper@{\NAT@date}}
  {}
  {}
\patchcmd{\NAT@citex}
  {\@citea\NAT@hyper@{%
     \NAT@nmfmt{\NAT@nm}%
     \hyper@natlinkbreak{\NAT@spacechar\NAT@@open\if*#1*\else#1\NAT@spacechar\fi}%
       {\@citeb\@extra@b@citeb}%
     \NAT@date}}
  {\@citea\NAT@nmfmt{\NAT@nm}%
   \NAT@spacechar\NAT@@open\if*#1*\else#1\NAT@spacechar\fi%
   \NAT@hyper@{\NAT@date}}
  {}
  {}
\makeatother

%% file: Latex_table.tex
 Magnitude & $c_{\rm gal}$ & $\log \rho_0$ & $\log \alpha$ & $\log r_{\rm s}$ &  $\log \rho_{\rm o}$ & $s_{\rm e}$ & $\log r_{\rm t}$ & $\log \beta$ & $\log \gamma$ & $R_{\rm sp}^{\rm 2d}$ & $R_{\rm sp}^{\rm 3d}$  & $\chi^2$/dof \\
\hline
-19.43 & high & $1.10_{-0.77}^{+0.25}$ & $-0.95_{-0.32}^{+0.22}$ & $-0.32_{-0.13}^{+0.40}$ & $0.349_{-0.035}^{+0.031}$ & $1.601_{-0.080}^{+0.076}$ & $-0.082_{-0.040}^{+0.049}$ & $0.762_{-0.095}^{+0.119}$ & $0.66_{-0.12}^{+0.14}$ & $0.778_{-0.014}^{+0.015}$ & $0.971_{-0.021}^{+0.025}$ & 6.0/8 \\
-19.43 & low & $-0.68_{-0.20}^{+0.30}$ & $-1.090_{-0.063}^{+0.088}$ & $0.55_{-0.17}^{+0.11}$ & $0.545_{-0.067}^{+0.055}$ & $1.600_{-0.080}^{+0.068}$ & $0.058_{-0.021}^{+0.023}$ & $1.10_{-0.11}^{+0.12}$ & $0.64_{-0.11}^{+0.13}$ & $1.153_{-0.021}^{+0.029}$ & $1.378_{-0.026}^{+0.026}$ & 13.2/8 \\
-20.43 & high & $0.70_{-0.86}^{+0.31}$ & $-0.97_{-0.35}^{+0.28}$ & $-0.27_{-0.16}^{+0.45}$ & $0.167_{-0.016}^{+0.014}$ & $1.613_{-0.077}^{+0.074}$ & $-0.098_{-0.038}^{+0.048}$ & $0.82_{-0.11}^{+0.13}$ & $0.69_{-0.12}^{+0.14}$ & $0.756_{-0.012}^{+0.014}$ & $0.938_{-0.026}^{+0.024}$ & 2.9/8 \\
-20.43 & low & $-0.89_{-0.28}^{+0.38}$ & $-1.019_{-0.086}^{+0.118}$ & $0.48_{-0.21}^{+0.15}$ & $0.276_{-0.021}^{+0.018}$ & $1.655_{-0.056}^{+0.052}$ & $0.072_{-0.023}^{+0.026}$ & $1.10_{-0.11}^{+0.12}$ & $0.80_{-0.13}^{+0.14}$ & $1.128_{-0.024}^{+0.029}$ & $1.352_{-0.025}^{+0.026}$ & 12.4/8 \\
-21.43 & high & $0.10_{-0.93}^{+0.39}$ & $-1.00_{-0.35}^{+0.33}$ & $-0.25_{-0.20}^{+0.49}$ & $0.0385_{-0.0048}^{+0.0045}$ & $1.496_{-0.099}^{+0.096}$ & $-0.087_{-0.044}^{+0.051}$ & $0.85_{-0.12}^{+0.14}$ & $0.78_{-0.13}^{+0.15}$ & $0.754_{-0.019}^{+0.022}$ & $0.938_{-0.040}^{+0.036}$ & 10.5/8 \\
-21.43 & low & $-1.31_{-0.37}^{+0.45}$ & $-0.97_{-0.12}^{+0.16}$ & $0.40_{-0.25}^{+0.20}$ & $0.0712_{-0.0066}^{+0.0063}$ & $1.624_{-0.073}^{+0.073}$ & $0.087_{-0.025}^{+0.026}$ & $1.12_{-0.13}^{+0.14}$ & $0.90_{-0.14}^{+0.15}$ & $1.132_{-0.035}^{+0.043}$ & $1.361_{-0.038}^{+0.034}$ & 23.6/8 

%% file: paper.bbl
\begin{thebibliography}{64}
\expandafter\ifx\csname natexlab\endcsname\relax\def\natexlab#1{#1}\fi

\bibitem[{{Adhikari} \& {Dalal}(2016)}]{Adhikari:2016}
{Adhikari}, S., \& {Dalal}, N. 2016, in preparation

\bibitem[{{Adhikari} {et~al.}(2014){Adhikari}, {Dalal}, \&
  {Chamberlain}}]{Adhikari:2014}
{Adhikari}, S., {Dalal}, N., \& {Chamberlain}, R.~T. 2014, \jcap, 11, 19

\bibitem[{{Aihara} {et~al.}(2011){Aihara}, {Allende Prieto}, {An}, {Anderson},
  {Aubourg}, {Balbinot}, {Beers}, {Berlind}, {Bickerton}, {Bizyaev}, {Blanton},
  {Bochanski}, {Bolton}, {Bovy}, {Brandt}, {Brinkmann}, {Brown}, {Brownstein},
  {Busca}, {Campbell}, {Carr}, {Chen}, {Chiappini}, {Comparat}, {Connolly},
  {Cortes}, {Croft}, {Cuesta}, {da Costa}, {Davenport}, {Dawson}, {Dhital},
  {Ealet}, {Ebelke}, {Edmondson}, {Eisenstein}, {Escoffier}, {Esposito},
  {Evans}, {Fan}, {Femen{\'{\i}}a Castell{\'a}}, {Font-Ribera}, {Frinchaboy},
  {Ge}, {Gillespie}, {Gilmore}, {Gonz{\'a}lez Hern{\'a}ndez}, {Gott}, {Gould},
  {Grebel}, {Gunn}, {Hamilton}, {Harding}, {Harris}, {Hawley}, {Hearty}, {Ho},
  {Hogg}, {Holtzman}, {Honscheid}, {Inada}, {Ivans}, {Jiang}, {Johnson},
  {Jordan}, {Jordan}, {Kazin}, {Kirkby}, {Klaene}, {Knapp}, {Kneib},
  {Kochanek}, {Koesterke}, {Kollmeier}, {Kron}, {Lampeitl}, {Lang}, {Le Goff},
  {Lee}, {Lin}, {Long}, {Loomis}, {Lucatello}, {Lundgren}, {Lupton}, {Ma},
  {MacDonald}, {Mahadevan}, {Maia}, {Makler}, {Malanushenko}, {Malanushenko},
  {Mandelbaum}, {Maraston}, {Margala}, {Masters}, {McBride}, {McGehee},
  {McGreer}, {M{\'e}nard}, {Miralda-Escud{\'e}}, {Morrison}, {Mullally},
  {Muna}, {Munn}, {Murayama}, {Myers}, {Naugle}, {Neto}, {Nguyen}, {Nichol},
  {O'Connell}, {Ogando}, {Olmstead}, {Oravetz}, {Padmanabhan},
  {Palanque-Delabrouille}, {Pan}, {Pandey}, {P{\^a}ris}, {Percival},
  {Petitjean}, {Pfaffenberger}, {Pforr}, {Phleps}, {Pichon}, {Pieri}, {Prada},
  {Price-Whelan}, {Raddick}, {Ramos}, {Reyl{\'e}}, {Rich}, {Richards}, {Rix},
  {Robin}, {Rocha-Pinto}, {Rockosi}, {Roe}, {Rollinde}, {Ross}, {Ross},
  {Rossetto}, {S{\'a}nchez}, {Sayres}, {Schlegel}, {Schlesinger}, {Schmidt},
  {Schneider}, {Sheldon}, {Shu}, {Simmerer}, {Simmons}, {Sivarani}, {Snedden},
  {Sobeck}, {Steinmetz}, {Strauss}, {Szalay}, {Tanaka}, {Thakar}, {Thomas},
  {Tinker}, {Tofflemire}, {Tojeiro}, {Tremonti}, {Vandenberg}, {Vargas
  Maga{\~n}a}, {Verde}, {Vogt}, {Wake}, {Wang}, {Weaver}, {Weinberg}, {White},
  {White}, {Yanny}, {Yasuda}, {Yeche}, \& {Zehavi}}]{Aihara:2011}
{Aihara}, H., {et~al.} 2011, \apjs, 193, 29

\bibitem[{{Bardeen} {et~al.}(1986){Bardeen}, {Bond}, {Kaiser}, \&
  {Szalay}}]{bardeen_etal86}
{Bardeen}, J.~M., {Bond}, J.~R., {Kaiser}, N., \& {Szalay}, A.~S. 1986, \apj,
  304, 15

\bibitem[{{Becker} \& {Kravtsov}(2011)}]{Becker:2011}
{Becker}, M.~R., \& {Kravtsov}, A.~V. 2011, \apj, 740, 25

\bibitem[{{Behroozi} {et~al.}(2013){Behroozi}, {Wechsler}, \&
  {Wu}}]{Behroozi:2013}
{Behroozi}, P.~S., {Wechsler}, R.~H., \& {Wu}, H.-Y. 2013, \apj, 762, 109

\bibitem[{{Bertschinger}(1985)}]{Bertschinger:1985}
{Bertschinger}, E. 1985, \apjs, 58, 39

\bibitem[{{Blanton} {et~al.}(2003){Blanton}, {Hogg}, {Bahcall}, {Brinkmann},
  {Britton}, {Connolly}, {Csabai}, {Fukugita}, {Loveday}, {Meiksin}, {Munn},
  {Nichol}, {Okamura}, {Quinn}, {Schneider}, {Shimasaku}, {Strauss}, {Tegmark},
  {Vogeley}, \& {Weinberg}}]{Blanton:2003}
{Blanton}, M.~R., {et~al.} 2003, \apj, 592, 819

\bibitem[{{Dalal} {et~al.}(2008){Dalal}, {White}, {Bond}, \&
  {Shirokov}}]{Dalal:2008}
{Dalal}, N., {White}, M., {Bond}, J.~R., \& {Shirokov}, A. 2008, \apj, 687, 12

\bibitem[{{Diemer} \& {Kravtsov}(2014)}]{Diemer:2014}
{Diemer}, B., \& {Kravtsov}, A.~V. 2014, \apj, 789, 1

\bibitem[{{Elbert} {et~al.}(2015){Elbert}, {Bullock}, {Garrison-Kimmel},
  {Rocha}, {O{\~n}orbe}, \& {Peter}}]{Elbert:2015}
{Elbert}, O.~D., {Bullock}, J.~S., {Garrison-Kimmel}, S., {Rocha}, M.,
  {O{\~n}orbe}, J., \& {Peter}, A.~H.~G. 2015, \mnras, 453, 29

\bibitem[{{Evans} \& {Bridle}(2009)}]{EvansBridle:2009}
{Evans}, A.~K.~D., \& {Bridle}, S. 2009, \apj, 695, 1446

\bibitem[{{Fillmore} \& {Goldreich}(1984)}]{fillmore_goldreich84}
{Fillmore}, J.~A., \& {Goldreich}, P. 1984, \apj, 281, 1

\bibitem[{{Foreman-Mackey} {et~al.}(2013){Foreman-Mackey}, {Hogg}, {Lang}, \&
  {Goodman}}]{Foreman-Mackey:2013}
{Foreman-Mackey}, D., {Hogg}, D.~W., {Lang}, D., \& {Goodman}, J. 2013, \pasp,
  125, 306

\bibitem[{{Gao} {et~al.}(2008){Gao}, {Navarro}, {Cole}, {Frenk}, {White},
  {Springel}, {Jenkins}, \& {Neto}}]{Gao:2008}
{Gao}, L., {Navarro}, J.~F., {Cole}, S., {Frenk}, C.~S., {White}, S.~D.~M.,
  {Springel}, V., {Jenkins}, A., \& {Neto}, A.~F. 2008, \mnras, 387, 536

\bibitem[{{Gao} {et~al.}(2005){Gao}, {Springel}, \& {White}}]{Gao:2005}
{Gao}, L., {Springel}, V., \& {White}, S.~D.~M. 2005, \mnras, 363, L66

\bibitem[{{Gao} \& {White}(2007)}]{Gao:2007}
{Gao}, L., \& {White}, S.~D.~M. 2007, \mnras, 377, L5

\bibitem[{{Gnedin} \& {Ostriker}(2001)}]{Gnedin:2001}
{Gnedin}, O.~Y., \& {Ostriker}, J.~P. 2001, \apj, 561, 61

\bibitem[{{Goodman} \& {Weare}(2010)}]{Goodman:2010}
{Goodman}, J., \& {Weare}, J. 2010, {Commun. Appl. Math. Comput. Sci.}, 5, 65

\bibitem[{{Harvey} {et~al.}(2015){Harvey}, {Massey}, {Kitching}, {Taylor}, \&
  {Tittley}}]{Harvey:2015}
{Harvey}, D., {Massey}, R., {Kitching}, T., {Taylor}, A., \& {Tittley}, E.
  2015, Science, 347, 1462

\bibitem[{{Hearin} {et~al.}(2014){Hearin}, {Watson}, \& {van den
  Bosch}}]{Hearin:2014}
{Hearin}, A.~P., {Watson}, D.~F., \& {van den Bosch}, F.~C. 2014, ArXiv
  e-prints

\bibitem[{{Hikage} {et~al.}(2013){Hikage}, {Mandelbaum}, {Takada}, \&
  {Spergel}}]{Hikage:2013}
{Hikage}, C., {Mandelbaum}, R., {Takada}, M., \& {Spergel}, D.~N. 2013, \mnras,
  435, 2345

\bibitem[{{Jiang} \& {van den Bosch}(2014)}]{Jiang:2014}
{Jiang}, F., \& {van den Bosch}, F.~C. 2014, ArXiv e-prints

\bibitem[{{Jiang} {et~al.}(2014){Jiang}, {Helly}, {Cole}, \&
  {Frenk}}]{Jiang:2014b}
{Jiang}, L., {Helly}, J.~C., {Cole}, S., \& {Frenk}, C.~S. 2014, \mnras, 440,
  2115

\bibitem[{{Kahlhoefer} {et~al.}(2014){Kahlhoefer}, {Schmidt-Hoberg},
  {Frandsen}, \& {Sarkar}}]{Kahlhoefer:2014}
{Kahlhoefer}, F., {Schmidt-Hoberg}, K., {Frandsen}, M.~T., \& {Sarkar}, S.
  2014, \mnras, 437, 2865

\bibitem[{{Kahlhoefer} {et~al.}(2015){Kahlhoefer}, {Schmidt-Hoberg}, {Kummer},
  \& {Sarkar}}]{Kahlhoefer:2015}
{Kahlhoefer}, F., {Schmidt-Hoberg}, K., {Kummer}, J., \& {Sarkar}, S. 2015,
  \mnras, 452, L54

\bibitem[{{Kaiser}(1984)}]{Kaiser:1984}
{Kaiser}, N. 1984, \apjl, 284, L9

\bibitem[{{Kang} {et~al.}(2007){Kang}, {van den Bosch}, {Yang}, {Mao}, {Mo},
  {Li}, \& {Jing}}]{Kang:2007}
{Kang}, X., {van den Bosch}, F.~C., {Yang}, X., {Mao}, S., {Mo}, H.~J., {Li},
  C., \& {Jing}, Y.~P. 2007, \mnras, 378, 1531

\bibitem[{{Klypin} {et~al.}(2014){Klypin}, {Yepes}, {Gottlober}, {Prada}, \&
  {Hess}}]{Klypin:2014}
{Klypin}, A., {Yepes}, G., {Gottlober}, S., {Prada}, F., \& {Hess}, S. 2014,
  ArXiv e-prints

\bibitem[{{Kravtsov} \& {Borgani}(2012)}]{Kravtsov:2012}
{Kravtsov}, A.~V., \& {Borgani}, S. 2012, \araa, 50, 353

\bibitem[{{Li} {et~al.}(2008){Li}, {Mo}, \& {Gao}}]{Li:2008}
{Li}, Y., {Mo}, H.~J., \& {Gao}, L. 2008, \mnras, 389, 1419

\bibitem[{{Lin} {et~al.}(2015){Lin}, {Mandelbaum}, {Huang}, {Huang}, {Dalal},
  {Diemer}, {Jian}, \& {Kravtsov}}]{Linetal:2015}
{Lin}, Y.-T., {Mandelbaum}, R., {Huang}, Y.-H., {Huang}, H.-J., {Dalal}, N.,
  {Diemer}, B., {Jian}, H.-Y., \& {Kravtsov}, A. 2015, ArXiv e-prints

\bibitem[{{Mandelbaum} {et~al.}(2005){Mandelbaum}, {Hirata}, {Seljak}, {Guzik},
  {Padmanabhan}, {Blake}, {Blanton}, {Lupton}, \&
  {Brinkmann}}]{Mandelbaum:2005}
{Mandelbaum}, R., {et~al.} 2005, \mnras, 361, 1287

\bibitem[{{Miyatake} {et~al.}(2015){Miyatake}, {More}, {Takada}, {Spergel},
  {Mandelbaum}, {Rykoff}, \& {Rozo}}]{Miyatake:2016}
{Miyatake}, H., {More}, S., {Takada}, M., {Spergel}, D.~N., {Mandelbaum}, R.,
  {Rykoff}, E.~S., \& {Rozo}, E. 2015, ArXiv e-prints

\bibitem[{{Mo} \& {White}(1996)}]{Mo:1996}
{Mo}, H.~J., \& {White}, S.~D.~M. 1996, \mnras, 282, 347

\bibitem[{{More}(2016{\natexlab{a}})}]{More:2016a}
{More}, S. 2016{\natexlab{a}}, in preparation

\bibitem[{{More}(2016{\natexlab{b}})}]{More:2016b}
---. 2016{\natexlab{b}}, in preparation

\bibitem[{{More} {et~al.}(2015){More}, {Diemer}, \& {Kravtsov}}]{More:2015}
{More}, S., {Diemer}, B., \& {Kravtsov}, A.~V. 2015, \apj, 810, 36

\bibitem[{{Navarro} {et~al.}(1996){Navarro}, {Frenk}, \&
  {White}}]{Navarro:1996ApJ}
{Navarro}, J.~F., {Frenk}, C.~S., \& {White}, S.~D.~M. 1996, \apj, 462, 563

\bibitem[{{Niikura} {et~al.}(2015){Niikura}, {Takada}, {Okabe}, {Martino}, \&
  {Takahashi}}]{Niikura:2015}
{Niikura}, H., {Takada}, M., {Okabe}, N., {Martino}, R., \& {Takahashi}, R.
  2015, \pasj

\bibitem[{{Oguri} {et~al.}(2012){Oguri}, {Bayliss}, {Dahle}, {Sharon},
  {Gladders}, {Natarajan}, {Hennawi}, \& {Koester}}]{Oguri:2012}
{Oguri}, M., {Bayliss}, M.~B., {Dahle}, H., {Sharon}, K., {Gladders}, M.~D.,
  {Natarajan}, P., {Hennawi}, J.~F., \& {Koester}, B.~P. 2012, \mnras, 420,
  3213

\bibitem[{{Oguri} \& {Hamana}(2011)}]{OguriHamana:2011}
{Oguri}, M., \& {Hamana}, T. 2011, \mnras, 414, 1851

\bibitem[{{Oguri} {et~al.}(2010){Oguri}, {Takada}, {Okabe}, \&
  {Smith}}]{Ogurietal:2010}
{Oguri}, M., {Takada}, M., {Okabe}, N., \& {Smith}, G.~P. 2010, \mnras, 405,
  2215

\bibitem[{{Oguri} {et~al.}(2005){Oguri}, {Takada}, {Umetsu}, \&
  {Broadhurst}}]{Oguri:2005}
{Oguri}, M., {Takada}, M., {Umetsu}, K., \& {Broadhurst}, T. 2005, \apj, 632,
  841

\bibitem[{{Patej} \& {Loeb}(2015)}]{Patej:2015}
{Patej}, A., \& {Loeb}, A. 2015, ArXiv e-prints

\bibitem[{{Randall} {et~al.}(2008){Randall}, {Markevitch}, {Clowe}, {Gonzalez},
  \& {Brada{\v c}}}]{Randall:2008}
{Randall}, S.~W., {Markevitch}, M., {Clowe}, D., {Gonzalez}, A.~H., \&
  {Brada{\v c}}, M. 2008, \apj, 679, 1173

\bibitem[{{Rines} {et~al.}(2013){Rines}, {Geller}, {Diaferio}, \&
  {Kurtz}}]{Rines:2013}
{Rines}, K., {Geller}, M.~J., {Diaferio}, A., \& {Kurtz}, M.~J. 2013, \apj,
  767, 15

\bibitem[{{Rozo} {et~al.}(2014){Rozo}, {Rykoff}, {Becker}, {Reddick}, \&
  {Wechsler}}]{Rozoetal:2014}
{Rozo}, E., {Rykoff}, E.~S., {Becker}, M., {Reddick}, R.~M., \& {Wechsler},
  R.~H. 2014, ArXiv e-prints

\bibitem[{{Rykoff} {et~al.}(2014){Rykoff}, {Rozo}, {Busha}, {Cunha},
  {Finoguenov}, {Evrard}, {Hao}, {Koester}, {Leauthaud}, {Nord},
  {et~al.}}]{Rykoff:2014}
{Rykoff}, E.~S., {et~al.} 2014, \apj, 785, 104

\bibitem[{{Rykoff} {et~al.}(2016){Rykoff}, {Rozo}, {Hollowood},
  {Bermeo-Hernandez}, {Jeltema}, {Mayers}, {Romer}, {Rooney}, {Saro}, {Vergara
  Cervantes}, {Wilcox}, {Abbott}, {Abdalla}, {Allam}, {Annis},
  {Benoit-L{\'e}vy}, {Bernstein}, {Bertin}, {Brooks}, {Burke}, {Capozzi},
  {Carnero Rosell}, {Carrasco Kind}, {Castander}, {Childress}, {Collins},
  {Cunha}, {D'Andrea}, {da Costa}, {Davis}, {Desai}, {Diehl}, {Dietrich},
  {Doel}, {Evrard}, {Finley}, {Flaugher}, {Fosalba}, {Frieman}, {Glazebrook},
  {Goldstein}, {Gruen}, {Gruendl}, {Gutierrez}, {Hilton}, {Honscheid}, {Hoyle},
  {James}, {Kay}, {Kuehn}, {Kuropatkin}, {Lahav}, {Lewis}, {Lidman}, {Lima},
  {Maia}, {Mann}, {Marshall}, {Martini}, {Melchior}, {Miller}, {Miquel},
  {Mohr}, {Nichol}, {Nord}, {Ogando}, {Plazas}, {Reil}, {Sahl{\'e}n},
  {Sanchez}, {Santiago}, {Scarpine}, {Schubnell}, {Sevilla-Noarbe}, {Smith},
  {Soares-Santos}, {Sobreira}, {Stott}, {Suchyta}, {Swanson}, {Tarle},
  {Thomas}, {Tucker}, {Viana}, {Vikram}, {Walker}, \& {Zhang}}]{Rykoff:2016}
---. 2016, ArXiv e-prints

\bibitem[{{Schlegel} {et~al.}(1998){Schlegel}, {Finkbeiner}, \&
  {Davis}}]{Schlegel:1998}
{Schlegel}, D.~J., {Finkbeiner}, D.~P., \& {Davis}, M. 1998, \apj, 500, 525

\bibitem[{{Sheth} {et~al.}(2001){Sheth}, {Mo}, \& {Tormen}}]{Sheth:2001}
{Sheth}, R.~K., {Mo}, H.~J., \& {Tormen}, G. 2001, \mnras, 323, 1

\bibitem[{{Sheth} \& {Tormen}(2004)}]{Sheth:2004}
{Sheth}, R.~K., \& {Tormen}, G. 2004, \mnras, 350, 1385

\bibitem[{{Skibba} {et~al.}(2011){Skibba}, {van den Bosch}, {Yang}, {More},
  {Mo}, \& {Fontanot}}]{Skibba:2011}
{Skibba}, R.~A., {van den Bosch}, F.~C., {Yang}, X., {More}, S., {Mo}, H., \&
  {Fontanot}, F. 2011, \mnras, 410, 417

\bibitem[{{Spergel} \& {Steinhardt}(2000)}]{Spergel:2000}
{Spergel}, D.~N., \& {Steinhardt}, P.~J. 2000, Physical Review Letters, 84,
  3760

\bibitem[{{Stoughton} {et~al.}(2002){Stoughton}, {Lupton}, {Bernardi},
  {Blanton}, {Burles}, {Castander}, {Connolly}, {Eisenstein}, {Frieman},
  {Hennessy}, {Hindsley}, {Ivezi{\'c}}, {Kent}, {Kunszt}, {Lee}, {Meiksin},
  {Munn}, {Newberg}, {Nichol}, {Nicinski}, {Pier}, {Richards}, {Richmond},
  {Schlegel}, {Smith}, {Strauss}, {SubbaRao}, {Szalay}, {Thakar}, {Tucker},
  {Vanden Berk}, {Yanny}, {Adelman}, {Anderson}, {Anderson}, {Annis},
  {Bahcall}, {Bakken}, {Bartelmann}, {Bastian}, {Bauer}, {Berman},
  {B{\"o}hringer}, {Boroski}, {Bracker}, {Briegel}, {Briggs}, {Brinkmann},
  {Brunner}, {Carey}, {Carr}, {Chen}, {Christian}, {Colestock}, {Crocker},
  {Csabai}, {Czarapata}, {Dalcanton}, {Davidsen}, {Davis}, {Dehnen},
  {Dodelson}, {Doi}, {Dombeck}, {Donahue}, {Ellman}, {Elms}, {Evans}, {Eyer},
  {Fan}, {Federwitz}, {Friedman}, {Fukugita}, {Gal}, {Gillespie}, {Glazebrook},
  {Gray}, {Grebel}, {Greenawalt}, {Greene}, {Gunn}, {de Haas}, {Haiman},
  {Haldeman}, {Hall}, {Hamabe}, {Hansen}, {Harris}, {Harris}, {Harvanek},
  {Hawley}, {Hayes}, {Heckman}, {Helmi}, {Henden}, {Hogan}, {Hogg}, {Holmgren},
  {Holtzman}, {Huang}, {Hull}, {Ichikawa}, {Ichikawa}, {Johnston}, {Kauffmann},
  {Kim}, {Kimball}, {Kinney}, {Klaene}, {Kleinman}, {Klypin}, {Knapp},
  {Korienek}, {Krolik}, {Kron}, {Krzesi{\'n}ski}, {Lamb}, {Leger},
  {Limmongkol}, {Lindenmeyer}, {Long}, {Loomis}, {Loveday}, {MacKinnon},
  {Mannery}, {Mantsch}, {Margon}, {McGehee}, {McKay}, {McLean}, {Menou},
  {Merelli}, {Mo}, {Monet}, {Nakamura}, {Narayanan}, {Nash}, {Neilsen},
  {Newman}, {Nitta}, {Odenkirchen}, {Okada}, {Okamura}, {Ostriker}, {Owen},
  {Pauls}, {Peoples}, {Peterson}, {Petravick}, {Pope}, {Pordes}, {Postman},
  {Prosapio}, {Quinn}, {Rechenmacher}, {Rivetta}, {Rix}, {Rockosi}, {Rosner},
  {Ruthmansdorfer}, {Sandford}, {Schneider}, {Scranton}, {Sekiguchi}, {Sergey},
  {Sheth}, {Shimasaku}, {Smee}, {Snedden}, {Stebbins}, {Stubbs}, {Szapudi},
  {Szkody}, {Szokoly}, {Tabachnik}, {Tsvetanov}, {Uomoto}, {Vogeley}, {Voges},
  {Waddell}, {Walterbos}, {Wang}, {Watanabe}, {Weinberg}, {White}, {White},
  {Wilhite}, {Wolfe}, {Yasuda}, {York}, {Zehavi}, \& {Zheng}}]{Stoughton:2002}
{Stoughton}, C., {et~al.} 2002, \aj, 123, 485

\bibitem[{{Tinker} {et~al.}(2012){Tinker}, {George}, {Leauthaud}, {Bundy},
  {Finoguenov}, {Massey}, {Rhodes}, \& {Wechsler}}]{Tinker:2012}
{Tinker}, J.~L., {George}, M.~R., {Leauthaud}, A., {Bundy}, K., {Finoguenov},
  A., {Massey}, R., {Rhodes}, J., \& {Wechsler}, R.~H. 2012, \apjl, 755, L5

\bibitem[{{Tinker} {et~al.}(2010){Tinker}, {Robertson}, {Kravtsov}, {Klypin},
  {Warren}, {Yepes}, \& {Gottl{\"o}ber}}]{Tinker:2010}
{Tinker}, J.~L., {Robertson}, B.~E., {Kravtsov}, A.~V., {Klypin}, A., {Warren},
  M.~S., {Yepes}, G., \& {Gottl{\"o}ber}, S. 2010, \apj, 724, 878

\bibitem[{{Tully}(2015)}]{Tully:2015}
{Tully}, R.~B. 2015, \aj, 149, 54

\bibitem[{{Umetsu} {et~al.}(2011){Umetsu}, {Broadhurst}, {Zitrin},
  {Medezinski}, \& {Hsu}}]{Umetsu:2011}
{Umetsu}, K., {Broadhurst}, T., {Zitrin}, A., {Medezinski}, E., \& {Hsu}, L.-Y.
  2011, \apj, 729, 127

\bibitem[{{Vogelsberger} {et~al.}(2011){Vogelsberger}, {Mohayaee}, \&
  {White}}]{vogelsberger_etal11}
{Vogelsberger}, M., {Mohayaee}, R., \& {White}, S.~D.~M. 2011, \mnras, 414,
  3044

\bibitem[{{Wechsler} {et~al.}(2006){Wechsler}, {Zentner}, {Bullock},
  {Kravtsov}, \& {Allgood}}]{Wechsler:2006}
{Wechsler}, R.~H., {Zentner}, A.~R., {Bullock}, J.~S., {Kravtsov}, A.~V., \&
  {Allgood}, B. 2006, \apj, 652, 71

\bibitem[{{Yang} {et~al.}(2006){Yang}, {Mo}, \& {van den Bosch}}]{Yang:2006}
{Yang}, X., {Mo}, H.~J., \& {van den Bosch}, F.~C. 2006, \apjl, 638, L55

\bibitem[{{Zentner} {et~al.}(2005){Zentner}, {Kravtsov}, {Gnedin}, \&
  {Klypin}}]{Zentner:2005}
{Zentner}, A.~R., {Kravtsov}, A.~V., {Gnedin}, O.~Y., \& {Klypin}, A.~A. 2005,
  \apj, 629, 219

\end{thebibliography}
